\documentclass[runningheads]{svmult}

\usepackage{graphicx}  
\usepackage{physmubb}  

\newcommand{\be}{\begin{equation}}
\newcommand{\ee}{\end{equation}}
\newcommand{\bea}{\begin{eqnarray}}
\newcommand{\eea}{\end{eqnarray}}

\newcommand{\ket}[1]{\left| #1 \right>}
\newcommand{\bra}[1]{\left< #1 \right|}
\begin{document}
\setcounter{chapter}{1}
\title{Basic Concepts and their Interpretation\footnote{Chapter 2 of
D. Giulini, E. Joos, C. Kiefer, J.
Kupsch, I.-O. Stamatescu, and H. D. Zeh: Decoherence and the Appearance
of a Classical World in Quantum Theory, 2nd edn. (Springer-Verlag,
2003). In this draft version, references to {\it other} chapters
refer to the first edition (Springer, 1996).} }
\toctitle{Basic Concepts and their Interpretation}
\titlerunning{Basic Concepts and their Interpretation}
\author{H.\ts D. Zeh \quad (www.zeh-hd.de)}
\authorrunning{H.\ts D. Zeh \quad (www.zeh-hd.de)}
\maketitle

\section    {The Phenomenon of Decoherence}
\subsection   {Superpositions}

The superposition principle forms the most fundamental kinematical concept
of quantum theory. Its universality seems to have first been
postulated by Dirac as part of the definition of his ``ket-vectors", which
he proposed as a complete\footnote{%
This {\it conceptual} completeness
does not, of course, imply that all degrees of freedom of a considered
system are always known and taken into account. It only means that,
within  quantum theory (which, in its way, is able to describe all known
experiments), no more complete  description of the system is required or
indicated. Quantum mechanics lets us even understand {\it why} we may
neglect certain degrees of freedom, since gaps in the energy spectrum
often ``freeze them out".  }
   and general concept to characterize
quantum states regardless of any {\it basis} of representation. They were
later recognized by von Neumann as  forming an abstract Hilbert space.
The inner product (also needed to  define a Hilbert space, and formally
indicated by the distinction between ``bra" and ``ket" vectors) is not
part of the kinematics proper, but required for the
probability  interpretation, which may be regarded as dynamics (as will
be discussed). The third Hilbert  space axiom (closure with respect to
Cauchy series) is merely mathematically	convenient, since
one can never decide empirically whether the number of linearly
independent physical states is infinite in reality, or just very large.

According to this kinematical superposition principle, {\it any} two
physical states,
$|  1\rangle$  and $|  2\rangle$, whatever their meaning, can be
superposed in the form $c_1 |  1\rangle + c_2 |  2\rangle$, with complex
numbers $c_1$ and
$c_2$, to form a {\it new physical state} (to be be distinguished from a
{\it state of information}).  By induction, the principle can be applied
to more than two, and even an infinite number, of states, and
appropriately generalized to apply to a continuum of states. After
postulating the linear Schr\"odinger equation in a general form, one may
furthermore conclude that the superposition of two (or more) of its
{\it solutions} forms again a solution. This is the {\it dynamical}
version of the superposition principle.

Let me emphasize that this superposition pinciple is in drastic contrast
to the concept of the ``quantum'' that gave the theory its name.
Superpositions obeying the Schr\"odinger equation describe a
deterministically evolving continuum rather than discrete
quanta and stochastic quantum jumps. According to the theory of
decoherence, these {\it effective} concepts ``emerge'' as a consequence
of the superposition principle when universally and consistently applied.

A dynamical superposition principle (though in general with respect
to real numbers only) is also known from classical waves which obey a
linear wave equation.  Its validity is then restricted to cases
where these equations apply, while the quantum superposition principle
is meant to be  universal and exact. However, while the physical meaning
of classical superpositions is usually obvious, that of quantum mechanical
superpositions has to be somehow determined. For example, the
interpretation of a superposition $\int dq\, e^{ipq} \, |q\rangle$ as
representing a state of momentum $p$ can be derived from
``quantization rules", valid for systems whose classical counterparts
are known in their Hamiltonian form (see Sect.\ts 2.2). In other cases, an
interpretation may be derived from the dynamics or has to be based on
experiments.

Dirac emphasized another (in his opinion
even more important)  difference: all
non-vanishing components of (or projections from) a superposition are ``in
some sense  contained" in it. This
formulation seems to refer to an  {\it ensemble} of physical states, which
would imply that their description by formal ``quantum
states" is {\it not} complete. Another interpretation
asserts that it is the (Schr\"
odinger) {\it dynamics} rather than the concept of quantum states which
is incomplete. States found in measurements would then have to {\it arise}
from an initial state by means of an indeterministic ``collapse of the
wave function". Both interpretations meet serious difficulties when
consistently applied (see Sect.\ts 2.3).

In the third edition of his textbook, Dirac  (1947) starts to explain the
superposition principle by discussing one-particle states, which
can be described by Schr\"odinger waves in three-dimensional space. This
is an important application, although its similarity with classical waves
may also be misleading. Wave functions derived from the quantization
rules are defined on their classical configuration space, which happens to
coincide with normal space only for a single mass point. Except for this
limitation, the two-slit interference experiment, for example,
(effectively a two-state superposition) is known to be very instructive.
Dirac's second example, the  superposition of two basic photon
polarizations, no longer corresponds to a spatial wave. These two basic
states ``contain" all possible photon polarizations. The electron spin,
another two-state system, exhausts the group  SU(2) by a two-valued
representation of spatial rotations, and it can be studied (with atoms or
neutrons) by means of many variations of the Stern--Gerlach experiment.
In his lecture notes (Feynman, Leighton, and Sands 1965), Feynman
describes the maser mode of the  ammonia molecule as another (very
different) two-state system.

All these examples  make essential use of superpositions of the kind $
\ket \alpha =  c_1\ket 1 + c_2 \ket 2$, where the states $\ket 1$,
$|  2\rangle$, and (all) $|  \alpha \rangle$ can be observed as
{\it physically different} states, and distinguished from one  another in
an appropriate setting. In the two-slit experiment, the states $|
1\rangle$  and $|  2\rangle$ represent the partial Schr\"odinger waves
that pass through one or the other slit. Schr\"odinger's wave function
can itself be  understood as a consequence of the superposition principle
by being viewed as the amplitudes $\psi_\alpha (q)$ in the superposition
of ``classical" configurations $q$ (now represented by corresponding
quantum states
$|  q\rangle$  or their narrow wave packets). In this
case of a system with a known classical counterpart, the superpositions
$| \alpha\rangle  = \int dq\,\psi_\alpha (q)|  q\rangle$ are assumed to
define all quantum states. They may represent new observable
properties (such as energy or angular momentum), which are
not simply functions of the configuration, $f(q)$, only as a nonlocal {\it
whole}, but not as an integral over  corresponding local densities
(neither  on space nor on configuration space).

Since Schr\"odinger's wave function is thus defined on (in general
high-dimensional) configuration space, increasing its amplitude
does not describe an increase of intensity or energy density, as
it would for classical waves in three-dimensional space. Superpositions
of the intuitive product states of composite quantum systems may
not only describe particle exchange symmetries (for bosons and fermions);
in the general case they  lead to the fundamental concept of {\it quantum
nonlocality}. The latter has to be distinguished from a mere {\it
extension} in space (characterizing extended classical objects). For
example, molecules in energy eigenstates are incompatible with their
atoms being in definite quantum states themselves. Although the
importance of this  ``entanglement" for many observable quantities (such
as the binding energy of the helium atom, or total angular momentum)
had been well known, its consequence of violating Bell's inequalities
(Bell 1964) seems to have surprised many physicists, since this result
strictly excluded all local theories conceivably underlying quantum
theory. However, quantum nonlocality appears paradoxical only when one
attempts to interpret  the wave function in terms of an ensemble of local
properties, such as ``particles".  If reality were {\it defined} to be
local (``in space and time"), then it would indeed conflict with the
empirical actuality of a general superposition. Within the quantum
formalism, entanglement also leads to decoherence, and in this way it {\it
explains} the classical appearance of the observed world in quantum
mechanical terms. The application of this program is the main subject of
this book (see also Zurek 1991, Mensky 2000, Tegmark and Wheeler 2001,
Zurek 2001, or www.decoherence.de).

The predictive power  of the superposition principle became particularly
evident when it was applied in an ingenious step to postulate the
existence of superpositions of states with different particle numbers
(Jordan and Klein 1927). Their meaning is illustrated, for example, by
``coherent states" of different photon numbers, which may represent
quasi-classical states of the electromagnetic field (cf. Glauber 1963).
Such dynamically arising (and in many cases  experimentally confirmed)
superpositions are often misinterpreted as representing ``virtual''
states, or mere probability amplitudes for the occurrence of ``real"
states that are assumed to possess definite particle number. This would
be as mistaken as replacing a hydrogen wave function by
the probability distribution
$p({\bf r}) = |\psi ({\bf r})|^2$, or an
entangled state by an {\it ensemble} of product states (or a two-point
function). A superposition is in general  observably different from an
ensemble consisting of its components with corresponding probabilities.

Another spectacular
success of the superposition  principle was the prediction of new
particles formed as superpositions of K-mesons and their antiparticles
(Gell-Mann and Pais 1955, Lee and Yang 1956). A similar model
describes the recently confirmed ``neutrino oscillations" (Wolfenstein
1978), which are superpositions of energy eigenstates.

The superposition principle can also be successfully applied to
states that may be generated by means of symmetry
transformations from asymmetric ones. In classical mechanics, a
symmetric Hamiltonian means that each asymmetric {\it solution} (such as
an elliptical Kepler orbit) implies other solutions, obtained
by applying the symmetry transformations (e.g. rotations). Quantum theory
requires {\it in addition} that all their superpositions  also form
solutions (cf. Wigner 1964, or Gross 1995;  see also Sect.\ts 9.6). A
complete set of energy eigenstates can then be constructed by means
of {\it irreducible linear representations}  of the dynamical symmetry
group. Among them are usually symmetric ones  (such
as s-waves for scalar particles) that need not have a  counterpart in
classical mechanics.

A great number of novel applications of the superpositon principle have
been studied experimentally or theoretically during recent years. For
example, superpositions of different ``classical'' states of laser modes
(``mesoscopic Schr\"odinger cats'') have been prepared (Monroe {\it et
al.} 1996), the entanglement of photon pairs has been confirmed to persist
over tens of kilometers (Tittel {\it et al.} 1998), and interference
experiments with fullerene molecules were successfully
performed (Arndt {\it et al.} 1999). Even superpositions of a macroscopic
current running in opposite directions have been shown to exist, and
confirmed to be different from a state with {\it two} (cancelling)
currents (Mooij {\it et al.} 1999, Friedman {\it et al.} 2000). Quantum
computers, now under intense investigation, would have to perform
``parallel" (but not {\it spatially} separated) calculations, while
forming one superposition that may later have a coherent effect.
So-called quantum teleportation requires the {\it advanced preparation}
of an entangled state of distant systems (cf. Busch et al. 2001 for a
consistent description in quantum mechanical terms). One of its
components may then later be selected by a {\it local} measurement in
order to determine the state of the other (distant) system.

Whenever an
experiment was technically feasible, all components of a superposition
have been shown to {\it act} coherently, thus proving that they {\it
exist} simultaneously. It is surprising that many physicists still seem to
regard superpositions as representing
some state of ignorance (merely characterizing unpredictable
``events''). After the fullerene experiments there remains but a minor
step to discuss conceivable (though hardly realizable)
interference experiments with a conscious observer.  Would he have one or
many ``minds'' (when being aware of his path through the slits)?

   The most general quantum  states seem to be
superpositions of different classical fields on three- or
higher-dimensional space.\footnote{The empirically correct ``pre-quantum''
configurations for fermions are given by spinor fields on space, while the
apparently observed particles are no more than the consequence of
decoherence by means of {\it local} interactions with the environment
(see Chap.\ts 3). Field amplitudes (such as
$\psi ({\bf r})$) seem to form the general arguments of the wave
function(al)
$\Psi$, while space points $\bf r$ appear as their ``indices'' -- not as
dynamical position variables.  Neither a ``second quantization'' nor a
wave-particle dualism are required.
$N$-particle wave functions may be obtained as a non-relativistic
approximation by applying the superposition principle (as a ``quantization
procedure'') to these apparent particles instead of the correct
pre-quantum variables (fields), which are not directly observable for
fermions. The concept of particle permutations then
becomes a {\it redundancy} (see Sect.\ts 9.6).  Unified field theories
are usually expected to provide a general  (supersymmetric) pre-quantum
field and its Hamiltonian.} In a perturbation expansion in terms of free
``particles" (wave modes) this leads to terms corresponding to Feynman
diagrams, as shown long ago by Dyson (1949). The path integral describes
a {\it superposition} of paths, that is, the propagation of wave
functionals according to a generalized Schr\"odinger equation, while the
individual paths under the integral have no {\it physical} meaning by
themselves. (A similar method could be used to describe the propagation
of classical waves.) Wave functions will here always be understood in the
generalized sense of {\it wave functionals} if required.

One has to keep in mind this universality of the superposition principle
and its consequences for {\it individually} observable physical properties
in order to appreciate the meaning of the program of decoherence.
Since quantum coherence is far more than the appearance of spatial
interference fringes observed statistically in series of ``events'',
decoherence must not simply be understood in a classical sense as their
washing out under  {\it fluctuating} environmental conditions.

\subsection   {Superselection Rules}
In spite of this success of the superposition principle it soon became
evident that not {\it all} conceivable superpositions are found in
Nature. This led some physicists to postulate ``superselection rules",
which restrict this principle by axiomatically excluding
certain superpositions (Wick, Wightman, and Wigner 1970, Streater
and Wightman 1964). There are also attempts to {\it derive} some of these
super\-selection rules from other principles, which	can be postulated
in quantum field theory (see Chaps. 6 and 7). In general, these
principles merely exclude ``unwanted" consequences of a general
superposition principle by hand.

Most disturbing in this sense seem to be superpositions of states with
integer and half-integer spin (bosons and fermions). They violate
invariance under
$2\pi$-rotations (see Sect.\ts 6.2), but such a non-invariance has
been experimentally confirmed in a different way (Rauch {\it et al.}
1975). The theory of supersymmetry (Wess and Zumino 1971) {\it postulates}
superpositions of bosons and fermions. Another supposedly
``fundamental" superselection rule forbids superpositions of
different charge.  For example, superpositions of a proton and a neutron
have never been directly observed, although they occur in the
{\it isotopic spin} formalism. This (dynamically broken) symmetry was
later successfully generalized to SU(3) and other groups in order to
characterize further intrinsic  degrees of freedom. However,
superpositions of a proton and a neutron may ``exist" within nuclei, where
isospin-dependent self-consistent potentials may arise from an {\it
intrinsic symmetry breaking}. Similarly, superpositions of different
charge are used to form BCS states (Bardeen, Cooper, and Schrieffer
1957), which describe the intrinsic properties of
superconductors. In these cases, definite
charge values have to be projected out (see Sect.\ts 9.6) in order to
describe the observed physical objects, which do obey the charge
superselection rule.

Other limitations of the superposition principle are less clearly defined.
While elementary particles are described by means of wave functions
(that is, superpositions of different positions or other properties), the
moon seems always to be at a definite place, and a cat is either dead or
alive. A general superposition principle would even allow superpositions
of a cat and a dog (as suggested by Joos). They would have to define a
"new animal" -- analogous to a $K_{long}$, which is a superposition of a
$K$-meson and its antiparticle. In the Copenhagen interpretation, this
difference is attributed to a strict conceptual separation between the
microscopic and the macroscopic world. However, where is the border line
that distinguishes an n-particle state of quantum mechanics from an
N-particle state that is classical? Where, precisely, does the
superposition principle break down?

Chemists do indeed know that a border line seems to exist
deep in the microscopic world  (Primas 1981, Woolley 1986). For example,
most molecules (save the smallest ones) are found	with their
nuclei in definite (usually rotating and/or vibrating) classical
``configurations", but hardly ever in superpositions thereof, as it would
be required for energy or  angular momentum eigenstates. The latter are
observed for hydrogen and other {\it small} molecules. Even chiral states
of a sugar molecule appear  ``classical", in contrast to its parity and
energy eigenstates, which correctly describe the otherwise
analogous maser mode states of the ammonia molecule (see Sect.\ts 3.2.4
for details). Does this difference mean  that quantum mechanics breaks
down already for very small particle number?

Certainly not in general, since there are well established
superpositions of many-particle  states: phonons in solids, superfluids,
SQUIDs, white dwarf stars and many more!  All properties of
macroscopic bodies which can be calculated quantitatively are consistent
with quantum mechanics, but not with any microscopic classical
description. As will be demonstrated throughout the book, the theory of
decoherence is able to {\it explain} the apparent differences between
the  quantum and the classical world under the assumption of a {\it
universally valid} quantum theory.

The attempt to derive the absence of certain superpositions
from (exact or approximate) conservation laws, which forbid or suppress
transitions between their corresponding components, would be insufficient.
This ``traditional" explanation (which seems to
be the origin of the name ``superselection rule") was used, for
example, by Hund (1927) in his arguments in favor of the chiral
states of molecules. However, small or vanishing transition rates require
{\it in addition} that superpositions were absent initially  for all these
molecules (or their constituents from which they formed). Similarly,
charge conservation does {\it not} explain the charge superselection rule!
Negligible wave packet dispersion (valid for large mass) may
prevent initially presumed wave packets from growing wider, but this
initial condition is quantitatively insufficient to explain the
quasi-classical appearance of mesoscopic objects, such as small dust
grains or large molecules (see Sect.\ts 3.2.1), or even that of celestial
bodies in chaotic motion (Zurek and Paz 1994).  Even initial
conditions for conserved quantities would in general allow one only to
exclude {\it global} superpositions, but not local ones (Giulini, Kiefer
and Zeh 1995).

So how can superselection rules be explained within quantum theory?

\subsection   {Decoherence by ``Measurements"}
Other experiments with quantum objects
have taught us that interference, for example between partial
waves, disappears when the property characterizing these partial waves
is {\it measured}. Such partial waves may describe the
passage through different slits of an interference device, or the two
beams of a Stern--Gerlach device (``{\it Welcher Weg} experiments"). This
loss of coherence is indeed  required by mere logic once
measurements are assumed to lead to definite results. In this case, the
frequencies of events on the detection screen measured in coincidence
with a {\it certain} passage can be counted separately, and thus have to
be added to define the total probabilities.\footnote{%
Mere
logic does {\it not} require, however, that the frequencies of events on
the screen which follow the observed passage through slit 1 of a two-slit
experiment, say, are the same as those without measurement, but with slit
2 closed. This distinction would be relevant in Bohm's theory  (Bohm
1952) if it allowed nondisturbing measurements  of the (now {\it assumed})
passage through one definite slit (as it does {\it not} in order to
remain indistinguishable from quantum theory). The fact that these two
quite different situations (closing  slit 2 or measuring the passage
through slit 1) lead to exactly the same subsequent frequencies, which
differ entirely from those that are {\it defined} by this theory when not
measured or selected,  emphasizes its extremely  artificial nature (see
also Englert {\it et al.} 1992, or Zeh 1999). The predictions of quantum
theory are here simply reproduced by leaving the Schr\"odinger equation
unaffected and universally valid, identical with Everett's
assumptions (Everett 1957). In both these theories the wave function is
(for good reasons) regarded as a {\it real} physical object (cf. Bell
1981).} It is
therefore a {\it plausible} experience that the interference disappears
also when the passage is ``measured" without  registration of a definite
result. The latter may be {\it assumed} to have become a ``classical
fact" as soon the measurement has irreversibly
``occurred". A quantum phenomenon may thus ``become a phenomenon'' {\it
without} being observed (in contrast to this early formulation of Bohr's,
which is in accordance with Heisenberg's idealistic statement about a
trajectory coming into being by its observation -- while Bohr later
spoke of objective irreversible events occurring in the counter). However,
what presicely is an irreversible quantum event? According to Bohr, it can
{\it not} be dynamically analyzed.

Analysis within the quantum mechanical formalism demonstrates nonetheless
that the essential condition for this ``decoherence" is that complete
information about the passage is carried
away in some {\it physical} form (Zeh 1970, 1973,
Mensky 1979, Zurek 1981, Caldeira and Leggett 1983, Joos and Zeh 1985).
Possessing  ``information" here means that the physical state of the
environment is now uniquely
{\it quantum correlated} (entangled) with the relevant property of the
system (such as a passage through a specific slit).
This need {\it not} happen in a controllable form (as in a measurement):
the ``information'' may as well be created in the form of noise. However,
in contrast to statistical correlations, quantum correlations define
{\it pure} (completly defined) nonlocal states, and thus
{\it individual physical} properties, such as the total spin of
spatially separated objects. Therefore, one cannot {\it
explain} entanglement in terms of the concept of information (cf.
Brukner and Zeilinger 2000). This terminology would mislead to the
popular misunderstanding of the collapse as a ``mere increase of
information'' (which would require an initial ensemble describing
ignorance). Since environmental decoherence affects individual physical
states, it can {\it neither} be the consequence of phase averaging in an
ensemble, {\it nor} one of phases fluctuating uncontrollably in time (as
claimed in some textbooks). For example, nonlocal entanglement exists in
the {\it static} quantum state of a relativstic physical vacuum (even
though it is then often visualized in terms of particles  as ``vacuum
fluctuations'').

When is unambiguous ``information" carried away?  If a macroscopic object
had the opportunity of passing through two slits, we would always be able
to convince ourselves of its choice of a path by simply opening our eyes
in order to ``look". This means that in this case there is plenty of
light that contains information about the path (even in a
controllable manner that allows ``looking''). Interference between
different paths never occurs, since the path is evidently ``continuously
measured" by light. The common textbook argument that the interference
pattern of macroscopic objects be too fine to be observable is entirely
irrelevant. However, would it then not be sufficient to dim the light
in  order to reproduce (in principle) a quantum mechanical interference
pattern for macroscopic objects?

This  could be investigated by means of more
sophisticated experiments with mesoscopic objects (see Brune {\it et al.}
1996). However, in order to precisely determine
the subtle limit where measurement by the environment becomes negligible,
it is  more economic first to apply the established theory which is known
to describe such experiments. Thereby we have to take into account the
quantum nature of the environment, as discussed long ago by Brillouin
(1962) for an information medium in general. This can usually be done
easily, since the quantum theory of  interacting systems, such as the
quantum theory of particle scattering, is well  understood. Its
application to decoherence requires that one averages over all unobserved
degrees of freedom. In technical terms, one has to ``trace out the
environment" after  it has interacted with the considered system. This
procedure leads to a quantitative theory of decoherence (cf. Joos and Zeh
1985). Taking the trace is based on the probability interpretation applied
to the environment (averaging over all possible outcomes of measurements),
even though this environment is {\it not} measured. (The precise physical
meaning of these formal concepts will be discussed in Sect.\ts 2.4.)

Is it possible to explain {\it all} superselection rules in this way as
an effect induced by the environment\footnote{It would be sufficient, for
this purpose, to use an {\it internal} ``environment'' (unobserved
degrees of freedom), but the assumption of a closed system would in
general be unrealistic.} -- including the existence and position of the
border line between microscopic and macroscopic behaviour in the realm of
molecules? This would mean that the universality of the superposition
principle could be maintained -- as is indeed the basic idea of the {\it
program of decoherence}  (Zeh 1970, Zurek 1982; see also Chap.\ts 4 of
Zeh 2001). If physical states are thus exclusively described by wave
functions rather than by points in configuration space -- as originally
intended by Schr\"odinger {\it in space} by means of narrow wave packets
instead of particles -- then no uncertainty relations are available {\it
for states} in order to explain the probabilistic aspects of quantum
theory: the Fourier theorem applies to a {\it given} wave function(al).

As another example, consider
two states of different  charge. They
interact very differently with the electromagnetic field
even in the  absence of radiation: their Coulomb fields carry complete
``information'' about the total charge  {\it at any distance}. The quantum
state of this field would thus decohere a superposition of different
charges if considered as a quantum system in a {\it bounded} region of
space  (Giulini, Kiefer, and Zeh 1995). This instantaneous action of
decoherence at an arbitrary distance by means of the Coulomb field gives
it the  appearance of a kinematical effect, although it is based on the
dynamical law of  charge conservation, compatible with a
{\it retarded} field that would ``measure'' the charge (see Sect.\ts
6.4).

There  are many other cases where the unavoidable effect of
decoherence can easily be imagined without any calculation. For example,
superpositions of  macroscopically
different electromagnetic fields, $f({\bf r})$, may be described by a
field functional $\Psi[f({\bf r})]$. However, any charged
particle in a  sufficiently narrow wave packet would then evolve into
different packets, depending on the field $f$, and thus become entangled
with the state of the quantum field (K\"ubler and Zeh 1973, Kiefer 1992,
Zurek, Habib, and Paz 1993; see also Sect.\ts 4.1.2). The particle  can
be said to ``measure" the quantum state of the field. Since charged
particles are in general abundant in the environment, no superpositions
of  macroscopically different electromagnetic fields (or different ``mean
fields'' in other cases) are observed under normal conditions. This result
is related to the difficulty of preparing and maintaining ``squeezed
states" of light (Yuen 1976) -- see Sect.~3.3.3.1. Therefore, the field
appears to {\it be} in one of its classical  states (Sect.~4.1.2).

In all these cases, this conclusion requires that the quasi-classical
states (or ``pointer states" in measurements) are robust (dynamically
stable) under natural decoherence, as pointed out already in the first
paper on decoherence (Zeh 1970; see also Di\'osi and Kiefer 2000).

A particularly  important example of a quasiclassical field is the metric
of general relativity (with classical {\it states} described by spatial
geometries on space-like hypersurfaces -- see  Sect.~4.2). Decoherence
caused by all kinds of matter can therefore explain the absence of
superpositions of macroscopically distinct spatial curvatures  (Joos
1986, Zeh 1986, 1988, Kiefer 1987), while {\it microscopic} superpositions
would describe those hardly ever observable gravitons.

Superselection  rules thus arise as a straightforward consequence of
quantum theory under realistic assumptions. They have nonetheless been
discussed mainly in mathematical physics -- apparently under the influence
of von Neumann's and Wigner's ``orthodox" interpretation of quantum
mechanics (see Wightman 1995 for a review).
Decoherence by ``continuous measurement" seems to form the most
fundamental irreversible process  in Nature. It applies even where
thermodynamical concepts do {\it not}  (such as for individual molecules
-- see Sect.~3.2.4), or when any exchange of heat is entirely
negligible. Its time arrow of ``microscopic causality" requires a
Sommerfeld radiation condition for microscopic scattering (similar to
Boltzmann's chaos), {\it viz.}, the absence of any dynamically relevant
{\it initial}  correlations, which would define a ``conspiracy" in common
terminology (Joos and Zeh 1985, Zeh 2001).
\vfil
\eject

\section    {Observables as a Derived Concept}
Measurements are usually described by means of ``observables", formally
represented by Hermitian operators,  and introduced in addition to the
concepts of quantum states and their dynamics as a  fundamental and
independent ingredient of quantum theory.  However, even though often
forming the starting point of a formal quantization procedure, this
ingredient should not be separately required if physical
states are well described by these formal quantum states. This
understanding, to be further explained below, complies with John Bell's
quest for the replacement of observables with ``beables" (see Bell 1987).
It was for this reason that his preference shifted from Bohm's theory to
collapse models (where wave functions are assumed to completely describe
{\it reality}) during his last years.

Let $|	\alpha\rangle$	be an arbitrary quantum state,  defined
operationally (up to a complex numerical factor) by a	``complete
preparation" procedure. The phenomenological probability for
finding the system during an appropriate measurement in another quantum
state
$|  n\rangle$, say, is given by means of their inner product as
$p_n = |\langle n\mid \alpha\rangle |^2$ (where both states  are
assumed to be normalized).  The state $|  n\rangle$
is here defined by the specific measurement. (In a position
measurement, for example, the number $n$ has to be replaced with the
continuous coordinates
$x,y,z$, leading to the ``improper" Hilbert  states $|\bf r \rangle$.) For
measurements of the ``first kind" (to which all others can be
approximately reduced -- see Sect.\ts 2.3),  the system will  again be
found in the state
$|  n\rangle$ with certainty if the measurement is  immediately repeated.
{\it Preparations} can be regarded as such measurements
which {\it select} a certain subset of outcomes for further measurements.
$n$-preparations are therefore also called $n$-filters, since all
``not-$n$" results are thereby excluded from the subsequent experiment
proper. The above probabilities can also be written in the form
$p_n = \langle \alpha\mid P_n\mid \alpha\rangle$, with  an
``observable" $P_n := |  n\rangle \langle n| $, which is thus {\it
derived} from the kinematical concept of quantum {\it states}.

Instead of these special ``$n$ or not-$n$ measurements" (with fixed $n$),
one can also  perform more general ``$n_1$ or $n_2$ or \dots\
measurements", with all
$n_i$'s mutually exclusive ($\langle n_i | n_j\rangle  =
\delta_{ij}$). If the states forming such a set $\{|  n\rangle \}$ are
pure and exhaustive (that is, complete, $\sum P_n = \bbbone$),
they represent a basis of the corresponding Hilbert space. By introducing
an arbitrary ``measurement scale" $a_n$, one may construct {\it general}
observables
$A = \sum |  n\rangle a_n\langle n|  $, which permit the definition of
``expectation values"  $\langle \alpha\mid A\mid \alpha\rangle  = \sum
p_na_n$. In the special case of a yes-no measurement, one has $a_n =
\delta_{nn_0}$, and expectation values become probabilities. Finding the
state
$ | n \rangle $ during a measurement is then also expressed as ``finding
the value $a_n$ of an observable". A
change of scale,
$b_n = f(a_n)$, describes the  {\it same} physical  measurement; for
position measurements of a particle it would simply represent a
coordinate transformation. Even a measurement of the particle's potential
energy
is equivalent to a position measurement (up to degeneracy) if the
function
$V(\bf r )$ is {\it given}.

According to this definition, quantum expectation values must not be
understood as mean values in an ensemble that represents
ignorance of the precise state. Rather, they have to be interpreted  as
probabilities for {\it potentially arising} quantum states
$|n\rangle$ -- regardless of the latters' interpretation.
If the set $\{|  n\rangle \}$ of such potential states forms a basis, any
state
$|  \alpha\rangle$  can be represented as a superposition
$|  \alpha\rangle  = \sum c_n|	n\rangle $. In general,  it neither forms
an $n_0$-state nor  any not-$n_0$ state. Its dependence on the complex
coefficients $c_n$ requires that states which differ from one another by a
numerical factor must be different ``in reality''. This is
true even though they represent the same ``ray" in Hilbert space and
cannot, according to the measurement postulate, be distinguished
operationally. The states
$|  n_1\rangle	+ |  n_2\rangle $ and $|  n_1\rangle   - |  n_2\rangle $
could not be physically different from another
   if $|	n_2\rangle $ and $-|  n_2\rangle $ were the {\it same}
state. (Only a {\it global}  numerical factor would be ``redundant''.)
For this reason, projection operators
$|n\rangle\langle n|$ are insufficient to characterize
quantum states (cf. also Mirman 1970).

The expansion coefficients $c_n$, relating physically meaningful states --
for example those describing different spin directions or different
versions of the K-meson -- must in principle be determined (relative to
one another) by appropriate experiments.  However, they can often be
derived from a previously known (or conjectured) classical  theory by
means of ``quantization rules". In this case, the classical
configurations $q$ (such as particle positions or field variables) are
{\it postulated} to parametrize a basis in Hilbert space,
$\{|  q\rangle \}$,  while the canonical momenta $p$ parametrize another
one,
$\{|  p\rangle \}$. Their corresponding observables,
$Q = \int dq\,|  q\rangle q\langle q| $  and
$P = \int dp\,|  p\rangle p\langle p| $,
are required to obey commutation relations in analogy to
the classical Poisson brackets. In this way, they form an important
{\it tool} for constructing and interpreting the specific Hilbert space of
quantum states. These commutators essentially determine the
unitary transformation
$\langle p\mid q\rangle $ (e.g. as a Fourier transform
$\E^{\I pq}$) -- thus more than what could be defined by means of the
projection operators $|q\rangle \langle q |$ and  $|p\rangle \langle p
|$. This algebraic procedure is mathematically  very elegant and
appealing, since the Poisson brackets and commutators may represent
generalized symmetry transformations. However, the {\it concept} of
observables (which form the algebra) can be derived from the more
fundamental one of state vectors and their inner products, as described
above.

Physical states are assumed to  vary in time in accordance with a
dynamical law -- in  quantum mechanics of the form
$\I \partial_t|  \alpha\rangle  = H|  \alpha\rangle $.  In contrast, a
measurement device is usually defined regardless of time. This must then
also hold for the observable representing it, or for its eigenbasis $\{
|  n\rangle \}$.  The probabilities
$p_n(t) = |\langle n\mid \alpha (t)\rangle |^2$ will therefore vary with
time according to the time-dependence of the  physical states $|
\alpha\rangle $. It is well known that this (Schr\"odinger) time
dependence is formally equivalent to the (inverse) time dependence of
observables (or the reference states
$|  n\rangle $). Since observables ``correspond" to classical  {\it
variables}, this time dependence appeared suggestive in the
Heisenberg--Born--Jordan algebraic approach to quantum theory.
However, the absence of {\it dynamical states} $|\alpha(t) \rangle$ from
this Heisenberg picture,  a consequence of insisting on {\it
classical} kinematical concepts, leads to paradoxes and
conceptual inconsistencies (complementarity, dualism, quantum logic,
quantum information, and all that).

An environment-induced superselection rule means that certain
superpositions are highly unstable with respect to decoherence.  It is
then impossible in practice to construct measurement devices for
them. This {\it empirical} situation has led some physicists to
{\it deny the existence} of these superpositions and their corresponding
observables --  either by postulate or by formal manipulations of dubious
interpretation, often including infinities. In an attempt to circumvent
the measurement problem (that will be discussed in the following section),
they often simply {\it regard} such superpositions as
``mixtures" once they have formed according to the Schr\"odinger equation
(cf. Primas 1990).

While {\it any} basis
$\{|  n\rangle \}$ in Hilbert space
defines formal probabilities,
$p_n =\break |\langle n | \alpha\rangle |^2$,  only a basis
consisting of states that are not immediately destroyed by decoherence
defines a practically ``realizable observable". Since realizable
observables usually form a genuine subset of {\it all} formal observables
(diagonalizable operators), they must contain a nontrivial ``center" in
algebraic terms. It consists of those of them which commute with all the
rest. Observables forming the center may be regarded as ``classical",
since they can be measured simultaneously with  {\it all} realizable
ones. In the algebraic approach to quantum theory, this center appears as
part of its axiomatic structure (Jauch 1968). However,
since the condition of decoherence has to be considered quantitatively
(and may even vary to some extent with the specific nature of the
environment), this  algebraic classification remains an approximate and
dynamically emerging scheme.

These ``classical" observables thus define the
subspaces into which superpositions decohere. Hence, even if the
superposition of a right-handed and a left-handed chiral molecule, say,
{\it could} be prepared by means of an appropriate (very fast)
measurement of the first kind, it would be destroyed
before the measurement may be repeated for a test. In contrast,
the chiral states of all individual molecules in a bag of sugar are
``robust" in a normal environment, and thus retain
this property {\it individually} over time intervals which by far exceed
thermal relaxation times. This stability may even be increased by the
quantum Zeno effect (Sect.~3.3.1). Therefore, chirality appears not only
classical, but also as an approximate constant of the motion that has to
be taken into account in the definition of thermodynamical ensembles (see
Sect.\ts 2.3).

The above-used description of measurements of the first kind by means
of probabilities for transitions
$|  \alpha\rangle  \to |  n\rangle $  (or, for that matter,
by corresponding observables) is phenomenological. However, measurements
should be described {\it dynamically} as interactions between the measured
system and the measurement device. The observable (that is, the
measurement basis) should thus be derived from the corresponding
interaction Hamiltonian and the initial state of the  device.
As discussed by von Neumann  (1932), this interaction must be
diagonal with respect to the  measurement basis
(see also Zurek 1981). Its diagonal matrix  elements are
operators which act on the quantum state of the device in such a way
that the  ``pointer" moves into a position appropriate for being read,
$|  n\rangle |	\Phi_0\rangle  \to |  n\rangle |  \Phi_n\rangle$. Here,
the first ket refers to the system, the second one to the device. The
states
$|  \Phi_n\rangle $, representing
different pointer positions, must approximately be mutually orthogonal,
and ``classical" in the explained sense.

Because of the dynamical superposition
principle, an initial
superposition
$\sum c_n|  n\rangle $ does {\it not} lead to definite pointer
positions (with their empirically observed  frequencies). If decoherence
is neglected, one obtains their {\it entangled superposition}
$\sum c_n|  n\rangle |	\Phi_n\rangle $, that is, a state that is
different from all potential measurement outcomes $|n\rangle |\Phi_n
\rangle$. This dilemma represents the ``quantum measurement problem"
to be discussed in Sect.\ts 2.3. Von Neumann's interaction is nonetheless
regarded as the first step of a measurement (a
``pre-measurement''). Yet, a collapse seems still to be  required -- now
in the measurement device rather than in the microscopic
system. Because of the entanglement between system and apparatus, it
then affects the total system.\footnote{%
    Some authors seem to
have taken the phenomenological collapse in the {\it microscopic  system}
by itself too  literally, and therefore disregarded the state of the
measurement device in their measurement theory (see Machida and Namiki
1980, Srinivas 1984, and Sect.\ts 9.1). Their approach is based
on the assumption that quantum states must always exist for all systems.
This would be in conflict with quantum nonlocality, even though it may be
in accordance with early interpretations of the quantum formalism.}

If, in a certain measurement, a whole subset of states $|n\rangle$ leads
to the same pointer position $| \Phi_{n_0} \rangle$, these
states are not distinguished in this measurement. The
pointer state
$|  \Phi_{n_0}\rangle $ now becomes dynamically
correlated with the whole {\it projection} of  the initial state,
$\sum c_n|  n\rangle $, on the subspace spanned by this
subset. A corresponding {\it collapse} was indeed postulated  by
L\"uders  (1951) in his generalization of von Neumann's ``first
intervention" (Sect.~2.3).

In this dynamical sense, the interaction with an appropriate measuring
device {\it defines} an observable up to arbitrary monotoneous scale
transformations. The time dependence of observables according to the
Heisenberg picture would thus describe an imaginary time dependence of
the states of this device (its pointer states), paradoxically controlled
by the intrinsic Hamiltonian of the {\it system}.

The  question of whether a formal observable (that is, a diagonalizable
operator) can be physically realized can only be  answered by
taking into account the unavoidable environment of the system (while the
measurement device is {\it always} asssumed to decohere into its
macroscopic pointer states). However, environment-induced decoherence by
itself does not solve the measurement problem, since the ``pointer
states" $|
\Phi_n\rangle$ may be assumed to {\it include} the total environment (the
``rest of the world").  Identifying the thus arising global
superposition with an {\it ensemble} of states, represented by a
statistical operator $\rho$, that merely leads to
the same {\it expectation values}  $\langle A\rangle = \rm{tr} (A\rho)$
for a {\it limited} set of observables $\{A\} $ would
obviously beg the question. This argument is nonetheless
found wide-spread in the literature (cf. Haag 1992, who used the
subset of all {\it local} observables).

In Sect.~2.4, statistical operators
$\rho$ will be {\it derived} from the concept of quantum states as a tool
for calculating expectation values, while the latter are defined, as
described above, by means of probabilities for the occurrence of new
states in measurements.  In the Heisenberg picture, $\rho$ is often
regarded as in some sense representing the {\it ensemble of potential
``values"} for all observables that are here postulated to formally
replace the classical variables. This interpretation is suggestive
because of the (incomplete) formal analogy of
$\rho$ to a classical phase space distribution. However, the prospective
``values" would be {\it physically} meaningful only if they characterized
different physical {\it states} (such as pointer states). Note that
Heisenberg's uncertainty relations refer to potential outcomes which may
arise in different (mutually exclusive) measurements.

\section    {The Measurement Problem}
The superposition of different measurement  outcomes, resulting
according to the Schr\"o\-dinger equation (as discussed above),
demonstrates that a ``naive  ensemble interpretation" of quantum
mechanics in terms of incomplete knowledge is ruled out. It would mean
that a  quantum state (such as
$\sum c_n|  n\rangle |	\Phi_n\rangle $) represents an  ensemble of some
as yet unspecified fundamental states, of which a subensemble (for
example  represented by the quantum state
$|  n\rangle |	\Phi_n\rangle $) may
be ``picked out by a mere increase of information". If this were
true, then the subensemble resulting from this measurement could in
principle be traced back in time by means of the Schr\"odinger
equation in order to determine also the initial state more completely
(to ``postselect" it -- see Aharonov and Vaidman 1991 for an inappropriate
attempt). In the above case this would lead to the initial
quantum state
$|  n\rangle |	\Phi_0\rangle $
that is {\it physically different} from -- and thus inconsistent with --
the superposition
$(\sum c_n| n\rangle )|
\Phi_0\rangle $ that had been prepared (whatever it {\it means}).

In spite of this simple argument,
which demonstrates  that an ensemble interpretation would  require a
complicated and miraculous nonlocal ``background  mechanism" in order to
work consistently (cf. Footnote 3 regarding Bohm's theory), the ensemble
interpretation of the wave function seems to  remain the most popular one
because of its pragmatic (though limited) value. A general and
rigorous critical discussion of problems arising in an ensemble
interpretation may be found in d'Espagnat's books (1976 and 1995).

A way out of this dilemma in terms of the wave function itself
requires one of the following two possibilities: (1) a modification of the
Schr\"odinger equation that explicitly describes a collapse (also called
``spontaneous localization" --
see Chap.~8), or (2) an Everett type interpretation, in which all
measurement outcomes  are assumed to coexist in one formal superposition,
but to be {\it perceived} separately as a  consequence of their dynamical
decoupling under decoherence. While this latter suggestion may appear
``extravagant" (as it requires myriads of coexisting parallel
quasi-classical ``worlds"), it is similar in principle to the conventional
(though nontrivial) assumption, made tacitly in all classical descriptions
of observations, that consciousness is {\it localized} in certain
(semi-stable and suffiently complex) {\it spatial subsystems} of the
world (such as human brains or parts thereof).  For a dispute about which
of the above-mentioned two possibilities should be preferred, the fact
that environmental decoherence readily describes precisely
the {\it apparently} occurring ``quantum jumps" or ``collapse events" (as
will be discussed  in great detail throughout this book) appears most
essential.

The dynamical rules which are (explicitly or tacitly)
used to describe the {\it effective} time dependence of quantum states
thus represent a ``dynamical dualism". This was
first clearly formulated by von Neumann  (1932), who distinguished
between the unitary evolution according to the Schr\"odinger equation
(remarkably his ``zweiter Eingriff" or ``second intervention"),
\be
      \I\hbar{\partial\over\partial t}\ket\psi = H\ket\psi \quad ,
      \ee
valid for isolated (absolutely closed) systems,  and the ``reduction" or
``collapse of the wave function",
\be
       \ket\psi = \sum c_n |  n\rangle  \to | n_0\rangle
       \ee
(remarkably his ``{\it first} intervention"). The latter was to describe
the stochastic transitions into the new state $|	n_0\rangle$  during
measurements.  Their dynamical discontinuity had been anticipated by
Bohr in the form of  ``quantum jumps" between his discrete electron
{\it orbits}. Later, the {\it time-dependent}
Schr\"odinger equation (2.1)  for interacting systems was often
regarded merely as a method of calculating probabilities for similar
(individually unpredictable) discontinuous transitions between
energy eigenstates (stationary {\it quantum} states) of atomic systems
(Born 1926).\footnote{%
Thus also Bohr
(1928) in a subsection entitled ``Quantum postulate and causality" about
``the quantum theory":  ``\dots its essence may be expressed in the
so-called quantum postulate,  which attributes to any {\it atomic
process} an essential discontinuity, or rather individuality,  completely
foreign to classical theories and symbolized by Planck's quantum of
action" (my italics). The  later revision of these early
interpretations of quantum theory (required by the important role of
entangled quantum states for much larger systems) seems to have	gone
unnoticed by many physicists.}
However, there does not seem to be any meaningful difference
between quantum jumps into new states or subspaces and
the ``occurrence of values" for certain observables (cf. Sect.~2.2).

In scattering theory, one usually probes only {\it part} of quantum
mechanics by restricting consideration to asymptotic states and their
probabilities (disregarding their superpositions). All quantum
correlations between them then appear statistical (``classical").
Occasionally even the unitary scattering
amplitudes
$\langle m_{out} | n_{in}
\rangle =
\bra{m} S
\ket{n}$ are confused with the {\it probability} amplitudes
$\langle \phi_m | \psi_n \rangle $ which describe measurements to find a
state $| \phi_m \rangle$ in an initial $|\psi_n \rangle$. In his general
S-matrix theory, Heisenberg temporarily speculated about deriving the
latter from the former. Since macroscopic systems never become asymptotic
because of their dynamical entanglment with the environment, they can not
be described by an S-matrix at all.

The Born/von Neumann dynamical dualism was evidently the major motivation
for an ignorance interpretation of the wave  function, which attempts to
explain the collapse {\it not} as a dynamical process in the system,  but
as an increase of {\it information} about it (the reduction of an
ensemble of {\it possible}  states). However, even though the dynamics of
ensembles in classical description uses a formally similar dualism, an
analogous interpretation in quantum theory leads to the severe (and
apparently fatal) difficulties indicated above. They are often
circumvented by the invention of ``new rules of logic and statistics",
which are {\it not} based on any ensemble interpretation or
incomplete information.

If the state of a {\it classical} system is incompletely known, and  the
corresponding point $p$,$q$
in phase space therefore replaced by an
ensemble (a probability distribution) $\rho (p,q)$, this
ensemble can be ``reduced" by a new observation that leads to increased
information. For this purpose, the system must interact
in a controllable manner with the ``observer" who holds the information
(cf. Szilard 1929). His physical state of memory must
thereby change in dependence on the property-to-be-measured of the
observed  system, leaving the latter unchanged in the ideal case (no
``recoil"). According to  {\it deterministic} dynamical laws, the
ensemble entropy of the combined system, which initially contains the
entropy corresponding to the unknown microscopic quantity, would remain
constant if it were defined to include the entropy characterizing
the final ensemble of different outcomes.  Since the observer is
assumed to ``know" (to be aware of) his own state, this  ensemble is
reduced correspondingly, and the ensemble entropy defined {\it with
respect to his state of information} is  lowered.

\begin{figure}[t]
\centering\includegraphics[width=.9\textwidth]{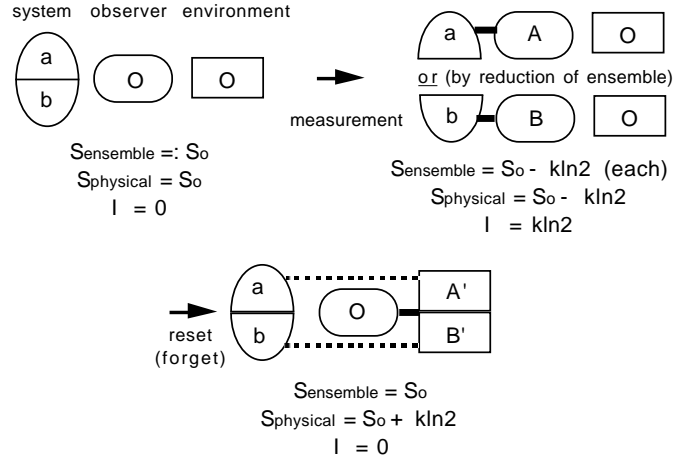}
\caption{{\footnotesize Entropy relative to the state of information in an
ideal {\it classical}
   measurement. Areas represent {\it sets} of microscopic
states of the subsystems (while those of uncorrelated combined systems
would be represented by their direct products).  During the first step of
the figure, the memory state of the observer changes deterministically
from $0$ to $A$ or
$B$, depending on the state $a$ or $b$ of the system to be measured.  The
second step depicts a subsequent reset,  required if the measurement is
to be repeated with the {\it same} device (Bennett 1973). $A'$ and $B'$
are  effects which must thereby arise in the thermal
environment in order to preserve the total ensemble entropy in accordance
with presumed microscopic determinism. The ``physical entropy"
({\it defined} to add for subsystems) measures the phase space of all
microscopic degrees  of freedom, including the property to be measured,
while {\it depending on given} macroscopic variables. Because of its
presumed additivity, this physical entropy neglects all remaining
statistical correlations (dashed lines, which indicate {\it sums} of
products of sets)  for being ``irrelevant" in the future --  hence
$S_{\rm physical} \geq S_{\rm ensemble}$. $I$ is the amount of information
held by the observer. The minimum initial entropy, $S_0$, is
$k\ln 2$ in this simple
case of two equally probable values $a$ and $b$.}}
\end{figure}

This is depicted by the first step of Fig. 2.1, where
ensembles of states are represented by areas. In contrast to many
descriptions of  Maxwell's demon, the observer (regarded as a device) is
here subsumed into the ensemble description.  {\it Physical
entropy}, unlike ensemble entropy, is usually understood as a {\it local}
(additive) concept, which neglects long range correlations for being
``irrelevant", and thus approximately  defines an {\it entropy density}.
Physical and ensemble entropy are equal in the absence of
correlations. The information
$I$, given in the figure, measures the reduction of entropy according to
the  increased knowledge of the observer.

This description
does not necessarily  require a {\it conscious} observer
(although it may ultimately rely upon him). It applies to any macroscopic
measurement device, since physical entropy is not only
defined to be local, but also relative to  ``given''  macroscopic
properties (as a function of them).  The dynamical part of the
measurement transforms
``physical" entropy (here the ensemble entropy of the microscopic
variables) deterministically into entropy of lacking information about
controllable macroscopic  properties. Before the observation is taken
into account (that is, before the ``or" is applied),  both parts of the
ensemble after the first step add up to give the ensemble entropy. When it
{\it is} taken into account (as done by the numbers given in the figure),
the ensemble entropy is  reduced according to the
information gained by the observer.

Any registration of information by the observer
must use up his memory capacity (``blank paper"),
which represents non-maximal entropy. If the
same measurement is to be repeated, for example in a cyclic process that
could be used to transform heat into mechanical energy (Szilard, {\it
l.c.}), this capacity would either be exhausted at some time, or an
equivalent amount of  entropy must be absorbed by the environment (for
example in the form of heat) in order to {\it reset} the measurement or
registration device (second step of Fig.~2.1). The reason is that two {\it
different} states cannot deterministically evolve into the same final
state (Bennett 1973).\footnote  {In {\it his} definitions, Bennett did
not count the entropy of the microscopic ensemble
$a$/$b$ as physical entropy, because this
variable is here assumed to be controllable, in contrast to the
``thermal" (ergodic or uncontrollable) property
$A'$/$B'$.}
This argument is based on an arrow of  time of ``causality'',
which requires that all correlations possess {\it local causes} in their
past (no ``conspiracy"). The irreversible formation of ``irrelevant"
correlations then explains the increase of {\it physical} (local)
entropy,  while the ensemble entropy is conserved.

\begin{figure}[t]
\centering\includegraphics[width=.9\textwidth]{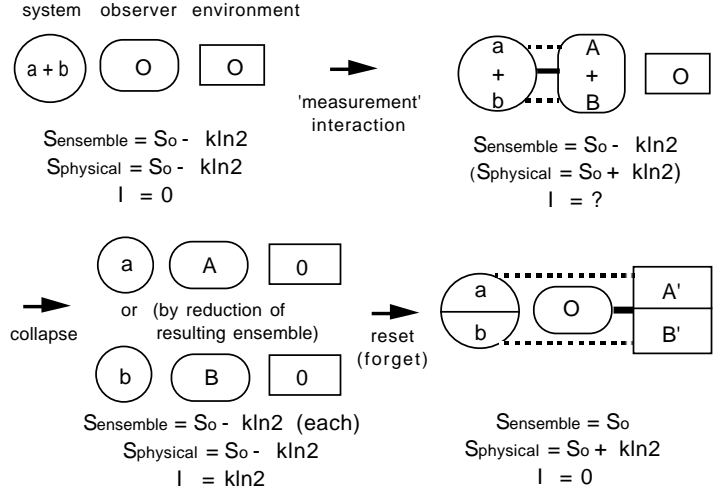}
\caption{{\footnotesize Quantum measurement of a {\it superposition}
$|  a\rangle  + |  b\rangle$ by means of a  {\it collapse} process, here
assumed to be triggered by the macroscopic pointer position. The initial
entropy is smaller by one bit than in Fig.~2.1 (and may in
principle vanish), since there is no initial {\it ensemble}  a/b for the
property to be measured. Dashed lines {\it before} the collapse now
represent quantum entanglement. (Compare the ensemble entropies
with those of Fig.\ts2.1!) Increase of physical entropy in the first step
is appropriate only if the arising entanglement is {\it regarded as
irrelevant}. The collapse itself is often  divided into {\it two} steps:
first increasing the ensemble entropy by replacing the superposition with
an ensemble, and then lowering it by reducing the ensemble (applying the
``or" -- for macroscopic pointers only). The increase of ensemble
entropy, observed in the final state of the Figure, is a consequence of
this  first step of the collapse. It brings the entropy up to its
classical initial value of Fig.\ts2.1}}
\end{figure}

The unsurmountable problems encountered in an ensemble
interpretation of the {\it wave function}
(or of any other superposition, such as
$|  a\rangle  + |  b\rangle $) are reflected by the fact that there is no
ensemble entropy
that would represent the unknown property-to-be-measured  (see the first
step of Fig. 2.2 or 2.3 -- cf. also Zurek 1984).
   The ``ensemble entropy" is now {\it defined} by the  ``corresponding"
expression
$S_{\rm ensemble} = -k\rm{tr} \{\rho\ln\rho\}$ (but see Sect.\ts 2.4 for
the meaning of the density matrix $\rho$). If the entropy of
observer plus environment is the same as  in the
classical case of Fig.\ts 2.1, the total initial ensemble entropy is now
lower; in the case of equal initial probabilities for $a$ and $b$ it is
$S_0 - k\ln 2$.
It would even vanish for pure states $\phi$ and $\chi$ of observer and
environment, respectively:
$(|  a\rangle +|  b\rangle )|  \phi_0\rangle |	\chi_0\rangle$.
The Schr\"odinger evolution (depicted in Fig. 2.3) would then be
described by three dynamical steps,
\bea
(\ket{a} +\ket{b})\ket{\phi_0} \ket{\chi_0}
    & \to& (\ket{a}\ket{\phi_A} +\ket{b} \ket{\phi_B}) \ket{\chi_0}
\nonumber
\\
   &\to& \ket{a} \ket{\phi_A} \ket{\chi_{A''}} + \ket{b}
\ket{\phi_B}
\ket{\chi_{B''}} \nonumber
\\
    & \to& (\ket{a} \ket{\chi_{A'A''}}+\ket{b}\ket{\chi_{B'B''}} )
\ket{\phi_0} \quad ,
  \eea
with an ``irrelevant"
(inaccessible)  final quantum correlation between system and environment
as a relic from the initial superposition. In this unitary evolution, the
   two
``branches" recombine to form a {\it nonlocal}
superposition. It ``exists, but it is not there''. Its local
unobservability characterizes an ``apparent collapse" (as will be
discussed). For a genuine collapse (Fig. 2.2), the final correlation
would be statistical, and the ensemble entropy would increase, too.

As mentioned in Sect.\ts 2.2, the general interaction  dynamics that is
required to describe ``ideal" measurements according to the Schr\"odinger
equation (2.1) is derived from the special case where the measured
system is prepared in an eigenstate $|n\rangle$ before measurement (von
Neumann 1932),
\be
      |  n\rangle  |  \Phi_0\rangle \to |  n\rangle |  \Phi_n\rangle \quad .
      \ee
Here,
$|  n\rangle$  corresponds to
$|  a\rangle$  or $|  b\rangle$  in the figures, the pointer
positions
$|  \Phi_n\rangle$ to the states
$|  \phi_A\rangle$  and $|  \phi_B\rangle $. (During non-ideal
measurements, the state
$| n\rangle$ would change, too.) However, applied to an initial
superposition,
$\sum c_n|  n\rangle $, the
interaction according to (2.1) leads to an entangled superposition,
\be
      \left( \sum c_n |  n\rangle \right) |  \Phi_0\rangle
      \to
      \sum c_n |	n\rangle    |  \Phi_n\rangle \quad .
      \ee
As explained in Sect.\ts 2.1.1, the resulting superposition represents an
{\it individual physical state} that is different from all
components  appearing in this sum. While
decoherence arguments teach us (see Chap.~3) that neglecting the
environment of (2.5) is absolutely unrealistic if $|
\Phi_n\rangle$ describes the pointer state of a macroscopic apparatus,
this superposition remains nonetheless valid if
$\Phi$ is defined to include
the ``rest of the universe", such as $|\Phi_n \rangle = |\phi_n \rangle
|\chi_n
\rangle$, with an environmental state $|\chi\rangle$. This powerful
consequence of the Schr\"odinger equation holds regardless of all
complications, such as decoherence and other, in practice irreversible,
processes (which need not  even be known). Therefore, it does seem that
the measurement problem can only be resolved if the Schr\"odinger
dynamics (2.1) is supplemented by a nonunitary  collapse (2.2).
 
\begin{figure}[t]
\centering\includegraphics[width=.9\textwidth]{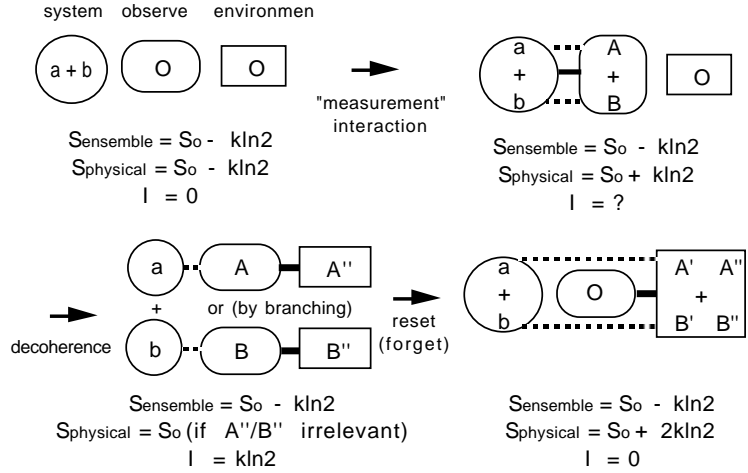}
\caption{{\footnotesize Quantum measurement of a superposition by
means of  ``branching" caused
by decoherence (see text). The
increase of physical entropy during the second step
applies if the distinction between environmental degrees of freedom
$A''/B''$, responsible for decoherence, is ``irrelevant"
(uncontrollable). After the last step, {\it all} entanglement has
irreversibly become irrelevant in practice. Since the whole superposition
is here assumed to ``exist" forever (and may have future consequences
{\it in principle}), the branching is  meaningful only with respect to a
{\it local} observer.}}
\end{figure}

Specific proposals for such a process will be
discussed in Chap.\ts 8.
Remarkably, however, there is no empirical evidence yet on where the
Schr\"odinger equation may have to be modified for this purpose
(see Joos 1986, Pearle and Squires 1994, or d'Espagnat 2001). On the
contrary, the dynamical superposition principle has been confirmed with
phantastic accuracy in spin systems (Weinberg 1989, Bollinger {\it et
al.} 1989).

The Copenhagen interpretation of quantum theory insists that the
measurement outcome has to be described in fundamental classical terms
rather than as a quantum state. While according to Pauli (in a letter to
Einstein: Born 1969), the appearance of an electron position
is ``a creation
outside of the laws of Nature" ({\it eine ausserhalb der Naturgesetze
stehende Sch\"opfung}), Ulfbeck and Bohr (2001) now claim (similar to
Ludwig 1990 in his attempt to derive ``the" Copenhagen interpretation from
fundamental principles) that it is the {\it click in the counter}
that appears ``out of the blue", and ``without an
event that takes place in the source itself as a precurser to the click''.
Together with the occurrence of this, thus {\it not dynamically
analyzable}, irreversible event in the counter, the wave function is then
claimed to ``lose its meaning" (precisely where it would otherwise
describe decoherence!). The Copenhagen interpretation is often hailed as
the greatest revolution in physics, since it rules out the
general applicability of the concept of objective physical reality.  I am
instead inclined to regard it as a kind of ``quantum voodoo":
irrationalism in place of dynamics. The theory of decoherence describes
events in the counter by means of a universal Schr\"odinger equation as a
fast and for all practical purposes irreversible {\it dynamical} creation
of entanglement with the environment (see also Shi 2000). In order to
remain ``politically correct'', some authors have recently even {\it
re-defined} complementarity in terms of entanglement (cf. Bertet {\it et
al.} 2001), although the latter has never been a crucial element of the
Copenhagen interpretation.

The
``Heisenberg cut" between observer and observed has often been claimed to
be quite arbitrary. This cut represents the borderline at which the
probability interpretation for the occurrence of events is applied.
However, shifting it too far into the  microscopic realm would miss the
readily observed quantum aspects of certain large systems (SQUIDs etc.),
while placing it beyond the detector would require the latter's
decoherence to be taken into account anyhow.
As pointed out by John Bell
(1981), the cut has to be placed  ``far enough" from  the
measured object in order to ensure that our limited capabilities of
investigation (such  as those of keeping the measured system isolated)
prevent us from discovering any inconsistencies with the assumed
classical properties or a collapse.

As noticed quite early in the historical debate, the cut
may even be placed deep into the human observer, whose consciousness,
which may be located in the cerebral cortex, represents the final
link in the observational chain. This view can be found in early
formulations by Heisenberg, it was favored by von Neumann, later
discussed by London and Bauer  (1939), and again supported by Wigner
(1962), among  others. It has even been interpreted as an {\it objective}
influence of consciousness on  physical reality  (e.g. Wigner {\it
l.c.}), although it may be consistent with the formalism only when used
with respect to {\it one} final observer, that is, in a strictly
subjective (though partly objectivizable) sense  (Zeh 1971).

The  ``undivisible chain between observer and
observed" is physically represented by a complex interacting medium, or a
chain of intermediary systems $|
\chi^{(i)}\rangle $, in quantum mechanical terms symbolically  written  as
\vfil
\eject
\noindent
\bea
      &&\ket{\psi_n^{\rm system}} \ket{\chi_0^{(1)}}\ket{\chi_0^{(2)}}
					      \cdots
			   \ket{\chi_0^{(K)}}\ket{\chi_0^{\rm obs}} 
\nonumber \\
    &\to&
       \ket{\psi_n^{\rm system}} \ket{\chi_n^{(1)}}\ket{\chi_0^{(2)}}
					      \cdots
				 \ket{\chi_0^{(K)}}\ket{\chi_0^{\rm 
obs}}    \nonumber  \\
     &\quad\! \vdots & \nonumber
							 \\
    &\to&
       \ket{\psi_n^{\rm system}} \ket{\chi_n^{(1)}}\ket{\chi_n^{(2)}}
					      \cdots
			       \ket{\chi_n^{(K)}}\ket{\chi_n^{\rm
obs}}  \quad ,
   \eea
instead of the simplified form (2.4).  This chain is thus
assumed to act dynamically step by  step  (cf. Zeh 1973). While  an
initial superposition of  the observed system now leads to a {\it
superposition} of such product states (similar to (2.5)), we know
empirically that a collapse must be  ``taken into account'' by the
conscious observer before (or at least when) the information arrives at
him as the final link. If there are several
chains connecting observer and observed (for example via  other
observers, known as ``Wigner's friends"), the correctly
applied Schr\"odinger equation warrants that each individual component
(2.6) describes consistent (``objectivized") measurement
results. From  the subjective point of view of the final observer, all
intermediary systems (``Wigner's friends" or ``Schr\"odinger's cats",
including their environments) could well remain in a superposition of
drastically different situations until he
observes (or communicates with) them!

Environment-induced decoherence means  that an avalanche of other causal
chains unavoidably branch off from the intermediary links of the chain as
soon as they become  macroscopic (see Chap.\ts 3). This might even
trigger a {\it real} collapse process (to be described by hypothetical
dynamical terms), since the	many-particle correlations arising from
decoherence would render the total system prone to such as yet
unobserved, but	nevertheless conceivable,  non-linear many-particle
forces (Pearle 1976, Di\'osi 1985, Ghirardi, Rimini, and Weber 1986,
Tessieri, Vitali, and Grigolini 1995;  see also Chap.~8). Decoherence by
a {\it microscopic} environment has been experimentally confirmed
to be {\it reversible} in what is now often called ``quantum
erasure" of a measurement (see Herzog et al. 1995).
In analogy to the concept of particle creation, reversible decoherence
may be regarded as ``virtual decoherence". ``Real" decoherence, which
gives rise to the familiar classical appearance of the macroscopic world,
is instead characterized by its unavoidability and irreversibility {\it in
practice}.

In an important contribution, Tegmark (2000) was able to demonstrate that
neuronal and other processes in the brain also become quasi-classical
because of environmental decoherence. (Successful neuronal models are
indeed classical.) This seems to imply that at least {\it objective}
aspects of human thinking and behavior can be described by conceptually
classical (though not necessarily deterministic) models of the brain.
However, since no precise ``localization of consciousness" within the
brain has been found yet, the neural network (just as the retina,
say) may still be part of the ``external world" with respect to the
unknown ultimate observer system (Zeh 1979).  Because of Tegmarks
arguments, this problem may not affect an {\it objective} theory of
observation any longer.

Even ``real" decoherence in the sense of above must be distinguished
from a genuine collapse, which is  defined as the {\it disappearance} of
all but one components from reality (thus representing an irreversible
{\it law}).\footnote{Proposed decoherence mechanisms involving event
horizons (Hawking 1987, Ellis, Mohanty and Nanopoulos 1989) would either
{\it require} a fundamental violation of unitarity, or merely represent a
specific kind of environmental decoherence (entanglement beyond the
horizon). The most immediate consequence of quantum entanglement is that
quantum theory can be consistently applied only to the whole universe.
      } As pointed out above, a  collapse could well occur much later
in the observational chain than decoherence, and possibly remain
less fine-grained. Nonetheless, it should then be detectable
in other situations if its dynamical rules are defined.
Environment-induced decoherence (the dynamically arising strong
correlations with  the rest of the world) leads to the important
consequence that, in  a world with no more than few-particle forces,
robust {\it factor} states $\ket{\chi_n^{\rm obs}}$ are not affected
by what goes on in the other branches that have formed according to the
Schr\"odinger equation.

In order to represent a subjective observer, a physical system must be in
a definite state with respect to properties of which he/she/it is
aware. The salvation of a psycho-physical  parallelism of this kind was
von Neumann's main argument for the introduction of his  ``first
intervention" (2.2): the collapse. As a consequence of the
above-discussed  dynamical independence of the different individual
components of type (2.6) in their  superposition, one may instead
associate {\it all} arising factor wave functions
$\ket{\psi_n^{\rm obs}}$
(different ones in each component) with {\it separate}  subjective
observers (that is, with different states of consciousness). This
approach, which avoids a collapse as a new dynamical law, is essentially
identical with Everett's ``relative state interpretation" (so called,
since the worlds {\it observed} by these observer states are described by
their corresponding relative factor states). Although also called a ``many
worlds interpretation", it describes {\it one} quantum universe. Because
of its (essential and non-trivial) reference to conscious observers, it
may more appropriately be called a ``multi-consciousness" or ``many minds
interpretation" (Zeh 1970, 1971, 1979, 1981, 2000, Albert and Loewer
1988, Lockwood 1989, Squires 1990, Stapp 1993, Donald 1995, Page
1995).\footnote{  As Bell (1981) pointed out, Bohm's theory would instead
require consciousness  to be psycho-physically coupled to certain {\it
classical} variables (which this theory {\it postulates} to exist). These
variables are probabilistically related  to the wave function by means of
a conserved statistical initial condition. Thus one may argue that
the ``many minds  interpretation" merely  eliminates Bohm's
unobservable and therefore meaningless intermediary classical variables
and their trajectories from this psycho-physical connection.
This is possible because of the  dynamical autonomy of a wave
function that evolves in time according to a universal Schr\"odinger
equation, and independently of Bohm's classical
variables. These variables thus cannot, by themselves, carry memories
of their ``surrealistic'' history. Memory is solely in the
quasi-classical wave packet that effectively guides them, while the other
myriads of ``empty" Everett world components (criticized for being
``extravagant" by Bell) {\it exist} as well in Bohm's  theory. Barbour
(1994, 1999), in his theory of timelessness, effectively proposed a {\it
static} version of Bohm`s theory, which  eliminates the latter's formal
classical trajectories even though it preserves a concept of memories
without a history (``time capsules" -- see also Chapt.\ts 6 of Zeh 2001).
						     }

Because of their dynamical
independence, all these different observers (or rather,
different ``versions" of the same observer) cannot find out by means of
experiments whether or not the other components have survived. This {\it
consequence of the Schr\"odinger equation}
thus leads to the  {\it impression} (for separate observers) that all
``other" components have ``hurried out of existence" as
soon as decoherence has become irreversible for all practical purposes.
Then it remains a pure matter of taste whether Occam's razor is applied to
the wave function (by adding appropriate but not directly detectable
collapse-producing nonlinear  terms
   to its dynamical law), or to the dynamical law (by instead
adding myriads of unobservable Everett components to our conception of
``reality").  Traditionally (and mostly successfully), consistency of the
law has been ranked higher than simplicity of the facts.

Fortunately,  the dynamics of decoherence can be discussed without
giving an answer to this question.  A collapse
(real or apparent)  has to be {\it taken into account} regardless of its
interpretation in order to describe the dynamics of that wave function
which represents our  {\it observed quasi-classical world} (the
time-dependent component  which contains ``our"
observer states
$\ket{\chi_n^{\rm obs}}$).  Only {\it specific}
dynamical collapse models could be confirmed or ruled out by experiments,
while Everett's relative states, on the other hand, depend {\it in
principle} on the definition of the observer system.\footnote{
Another aspect of this
observer-relatedness of the observed world is the concept of a {\it
presence}, which is not part of {\it physical}
time. It reflects the {\it empirical fact} that the subjective
observer is local in space {\it and} time.} (No other ``systems" have to
be specified in principle. Their density matrices, which describe
decoherence and quasi-classical concepts, are merely convenient.)

In contrast to Bohm's theory or stochastic collapse models, nothing has
been said yet (or postulated) about {\it probabilities} of measurement
outcomes. For this purpose, the Everett branches have to be given
statistical weights in a way that appears {\it ad hoc} again.
However, these probabilities are meaningful to an observer only as
frequencies in {\it series} of equivalent measurements. These measurements
must be performed in his branch (and would in general be performed and
lead to different results in other branches).  Graham (1970) was able to
show that the norm of the superposition of {\it all those} Everett
branches (arising from such series of measurements) which contain
frequencies of results that substantially {\it differ} from Born's
probabilities vanishes in the limit of infinite series. Although the
definition of the norm, which is used in this argument, is
exactly equivalent to Born's probabilties, it can be selected against
other definitions of a norm by its unique property of being
conserved under the Schr\"odinger dynamics.

To give an example: an isolated decaying quantum system may be described
as a superposition of its metastable initial state and the outgoing
channel wave function(s) for all its decay  products. On a large but
finite region of space and time the total wave function may then {\it
approximately} decrease exponentially and coherently in accordance with
the Schr\"odinger equation (with very small late-time deviations from
exponential behavior caused by the dispersion of the outgoing waves). For
a system that decays by emitting photons into a reflecting cavity, a {\it
superposition of different decay times} has in fact been confirmed in the
form of coherent ``state vector revival" (Rempe, Walther and Klein 1987).
An even more complex experiment exhibiting coherent state vector revival
was performed by means of spin waves (Rhim,
Pines and Waugh 1971). In  general, however, the decay fragments would
soon be ``monitored" by surrounding matter. The  resulting state of the
environment must then depend (contain ``information") on the  decay time,
and the superposition will decohere into dynamically independent
components corresponding to {\it different}  (approximately defined)
decay times. From the point of  view of a local observer, the system may
be assumed to {\it have decayed at a certain time} (within the usually
very narrow limits set by the decoherence time scale -- see Sect.~3.3.2),
even though he need not have observed the decay. This  situation does not
allow coherent state vector revival any more. Instead, it leads to an
exponential distribution of decay times in the arising apparent ensemble,
valid shortly after the decaying state has been produced (see Joos 1984).

However, as long as the information has not yet reached the observer,
\be
      \biggl( \sum_n c_n \ket{\psi_n^{\rm system}}
\ket{\chi_n^{(1)}}\ket{\chi_n^{(2)}}
\cdots		       \ket{\chi_n^{(K)}} \biggr) \ket{\chi_0^{\rm obs}},
    \ee
he may as well assume  (from his subjective point of view) that the
nonlocal superposition still exists.  According to the formalism,
Schr\"odinger's cat (represented by
$\ket{\chi^{(2)}}$, say) would then
``become" dead {\it or} alive only when he becomes aware of it.  On the
other hand, the property described by the state
$\ket{\psi_n^{\rm system}}$ (just as the cat's status of being dead or
alive,
$\ket{\chi_n^{(2)}}$) can also be assumed to have become ``real" as soon
as decoherence has become irreversible in practice. Therefore, decoherence
must also corrupt any controllable entanglement that would give rise to a
violation of Bell's inequalities (as it does -- see Venugopalan, Kumar
and Gosh 1995). If, instead of taking notice  of the result, the  observer
would decide to perform another measurement of the ``system"  (by using
a new observational chain), he could not observe any interference between
different n's, since, as a local  system, he cannot perform the required
global measurements. All predictions which this observer  can check are
consistent with the assumption that the system	was in {\it one} of
the states
$\ket{\psi_n^{\rm system}}$ (with probability $|c_n|^2$) before the second
measurement (see also Sect.~2.4).  This justifies the interpretation that
the cat is {\it determined} to die or not yet to die as soon as
irreversible decoherence has occurred somewhere in the chain (which will
in general be the case {\it before} the poison is applied).


\section    {Density Matrix, Coarse Graining, and
``Events"}
The theory of decoherence uses some (more or less technical) auxiliary
concepts. Their physical meaning will be recalled and discussed in this
section, as it is essential for a correct interpretation of what is
actually achieved with this theory.

    In classical
statistical mechanics,
{\it incomplete knowledge} about the real physical state of a system is
described by ``ensembles'' of states, that is, by probability
distributions
$\rho(p,q)$ on phase
space (in Fig.\ts 2.1 symbolically indicated by  areas of uniform
probability). Such ensembles are often called ``thermodynamic'' or
``macroscopic states''. Mean values of state functions
$a(p,q)$ (that is, physical quantities that are determined by the
microscopic state p,q), defined with respect to this ensemble, are then
given by the expression
$\int dp\, dq\, \rho(p,q) a(p,q)$. The ensemble $\rho (p,q)$ itself could
be recovered from the mean values of a complete set of state functions
(such as all $\delta$-functions), while a (smaller) set, that may be
realized in practice, determines only a ``coarse-grained" probability
distribution.

{\it If} all states which form such an ensemble are assumed to obey the
same Hamiltonian equations, their probability distribution $\rho$ evolves
according to the Liouville equation,
\be
      {\partial\rho\over\partial t} = \left\{H \mathbin, \rho \right\} \quad ,
      \ee
with Hamiltonian $H$ and Poisson bracket $\{,\}$.
However, this assumption
would be highly unrealistic for a many-particle system.  Even if the {\it
fundamental} dynamics is assumed to be given, the
{\it effective} Hamiltonian for the considered system depends
very sensitively on the state of the ``environment",  which cannot
be assumed  to be known better than that of the system
itself. Borel  (1914) showed long ago that even the gravitational effect
resulting from shifting a small rock at the distance of Sirius by a
few centimetres would completely change the microscopic state of a	gas in
a vessel here on earth within seconds after the retarded field
has arrived (see also Chap.~\ts 3). In a similar connection, Ernst Mach
spoke of the ``profound interconnectedness of things". This  surprising
result is facilitated by the enormous amplification of the tiny
differences in the molecular trajectories, caused by the slightly
different forces, during subsequent collisions with other molecules (each
time by a factor of the order of the ratio  of the mean free path over
the molecular radius). Similarly, microscopic differences in the state of
the gas will immediately disturb its environment, thus leading in turn
to  slightly different effective Hamiltonians for the gas, with in
general grossly different (``chaotic") effects on the microscopic states
of the original ensemble. This will induce strong  {\it statistical}
correlations of the gas with its environment, whose neglect would
lead to an increase of ensemble entropy.

This {\it effective local dynamical indeterminism} can be
taken into account (when calculating {\it forward} in time) by
means of stochastic forces (using a {\it Langevin equation}) for the
individual state, or by means of a corresponding {\it master equation} for
an ensemble of states,
$\rho (p,q)$. The increase of the local ensemble
entropy is thus attributed to an uncertain effective Hamiltonian. In this
way, statistical correlations with the environment are regarded as
dynamically irrelevant for the {\it future} evolution. An example is
Boltzmann's collision equation (where the arising irrelevant
correlations are {\it intrinsic} to the gas, however). The
justification of this time-asymmetric procedure forms a basic problem of
physics and cosmology (Zeh 2001).

When appying the conventional quantization rules to the
Liouville equation (2.8) in a formal way, one obtains the {\it von
Neumann equation} (or {\it quantum Liouville equation}),
\be
      \I\hbar {\partial\rho\over\partial t} = \left[ H \mathbin, \rho
\right] \quad
,
      \ee
for the dynamics of ``statistical operators" or ``density operators"
$\rho$. Similarly, expectation values
$\langle A\rangle  = \rm{tr}(A\rho)/\rm{tr}(\rho)$ of observables $A$
formally replace
   mean values $\bar a = \int dp\, dq\, a(p,q)\rho (p,q)$ of the state
functions
$a(p,q)$. Expectation values of a {\it restricted}  set of observables
would again represent a generalized coarse graining for the density
operators.  The  von Neumann equation (2.9) is unrealistic
for similar reasons as is the Liouville equation,
although quantitative differences may arise from  the different energy
spectra -- mainly at low energies. Discrete spectra
have relevant consequences for  macroscopic systems in practice only in
exceptional cases, while they often prevent mathematically
rigorous proofs of ergodic or chaos theorems which are valid in
excellent approximation. However, whenever quantum
correlations do form in analogy to classical correlations (as is the
rule), they lead to far more profound consequences than  their classical
counterparts.

In order to explain these differences, the concept of a density
matrix has to be derived from that of a (pure) quantum state instead of
being postulated by means of quantization rules. According to Sect.\ts
2.2, the probability for a state
$|  n\rangle$  to be ``found'' in a state $|  \alpha\rangle$	in a
corresponding measurement is given by
$|\langle n\mid \alpha\rangle |^2$. Its {\it mean}
probability in an  ensemble of states $\{ |  \alpha\rangle \}$
with probabilities
$p_\alpha$ representing incomplete information about the initial state
$\alpha$ is, therefore,
$p_n = \sum p_\alpha|\langle n\mid \alpha\rangle |^2 = \rm{tr}\{\rho
P_n\}$, where
$\rho = \sum |	\alpha\rangle p_\alpha\langle \alpha| $  and
$P_n = |  n\rangle \langle n|  $. This result remains true for general
observables
$A = \sum a_nP_n$ in place of
$P_n$. The ensemble of wave functions
$|  \alpha\rangle $, which thus defines a density matrix as representing
a {\it state of information}, need not consist of mutually orthogonal
states, although the density matrix can always be diagonalized in terms
of its eigenbasis. Its representation by a {\it general} ensemble of
states is therefore far from unique -- in contrast  to a classical
probability distribution. Nonetheless, the density matrix can still be
shown to obey a von Neumann equation {\it if} all states contained in the
ensemble are  assumed to evolve according to the same unique
Hamiltonian.

However,  the fundamental nonlocality of quantum states means that the
state of a local system does {\it not exist} in general: it cannot be
merely unknown. Accordingly, there {\it is no} effective local
Hamiltonian that would allow (2.9) to apply in principle (see K\"ubler
and Zeh 1973). In particular, a time-dependent Hamiltonian would in
general require a (quasi-)classical environment. This specific quantum
aspect is easily overlooked when the density matrix is introduced
axiomatically by ``quantizing" a classical probability distribution on
phase space.

Quantum nonlocality means that the generic state of a composite
system (``system" and environment, say),
\be
      \ket\Psi = \sum_{m,n} c_{mn} \ket{\phi_m^{\rm system}}
						    \ket{\phi_n^{\rm
environment}} \quad ,
      \ee
does not factorize. The expectation values of all {\it local} observables,
\be
      A = A^{\rm system} \otimes \bbbone^{\rm environment} \quad ,
      \ee
have then to be calculated by ``tracing out" the environment,
\be
      \bra\Psi A \ket\Psi \equiv \rm{tr} \left\{ A \ket\Psi \bra\Psi \right\}
= {\rm trace}_{\rm system} \left\{ A^{\rm system}
\rho^{\rm system}
\right\} \quad .
      \ee
Here, the density matrix $\rho^{\rm system}$, which in general has nonzero
entropy even for a pure (completely defined) global state $|
\Psi
\rangle$, is given by
\bea
	\rho^{\rm system}
	 \equiv
	\sum_{m,m'} \ket{\phi_m^{\rm system}}
	\rho_{mm'}^{\rm system}\bra{\phi_{m'}^{\rm system}}    \nonumber    \\
	: =
	 {\rm trace}_{\rm env} \left\{ \ket\Psi \bra\Psi \right\}
	\equiv \sum_{m,m'} \ket{\phi_m^{\rm system}}
		  \sum_n c_{mn} c_{m'n}^\ast
		    \bra{\phi_{m'}^{\rm system}}.
    \eea
It represents the specific
{\it coarse-graining} with respect to all {\it subsystem}
observables only. This ``reduced density matrix" can be
{\it formally} represented by various
ensembles of local states (including its eigenrepresentation or diagonal
form), although it does here characterize one pure but
entangled {\it global} state.  A density matrix thus based on entanglement
has been called an ``improper mixture" by d'Espagnat (1966). It can
evidently not explain {\it ensembles} of definite measurement outcomes.
If proper and improper mixtures were identified for
operationalistic reasons (that are {\it based on} the measurement
postulate), then decoherence would indeed completely ``solve'' the
measurement problem.

Regardless of its origin and interpretation, the density matrix can
be replaced by its partial Fourier transform, known as the
{\it Wigner function} (see also Sect.\ts 3.2.3):
\eject
\noindent
\bea
W(p,q):&=& {1 \over \pi} \int {e^{2\I px} \rho(q+x,q-x)\, dx} \nonumber\\
&\equiv&  {1\over {2\pi}}
\int \int \delta \left( q-{{z+z^\prime}\over 2}\right)
\E^{\I p(z-z^\prime)}
\rho(z,z^\prime)
\, dz dz^\prime   \nonumber
\\
  &=&: {\rm trace} \{ \Sigma_{p,q} \rho \} \quad .
\eea
The second line is here written in analogy to the Bloch vector, ${
\pi_i} = {\rm trace}\{ {\bf \sigma_i}\rho\}$, since
\be
\Sigma_{p,q}(z,z^\prime)  :=
{1 \over {2\pi}} e^{\I p(z-\nobreak z^\prime)}\, \delta \left( q-\nobreak
{z+z^\prime\over 2}\right) \quad  \ee
   is a generalization of the Pauli matrices
(with index $p,q$ instead of $i=1,2,3$ -- see also Sect.\ts 4.4
of Zeh 2001).  Although the Wigner function is {\it formally} analogous
to a phase space distribution, it does, according to its derivation,
{\it not} represent an ensemble of classical states (phase space points).
This is reflected by its potentially negative values, while even a
Gaussian wave packet, which does lead to a non-negative Wigner function,
is nonetheless {\it one} (pure) quantum state.

The degree of entanglement represented by an improper mixture (2.13) is
conveniently measured by the latter's formal entropy, such as
the linear entropy $S_{lin} = {\rm trace} (\rho - \rho^2)$. In a
``bipartite system", the {\it mutual} entanglement of its two
parts may often be controlled and {\it used} for specific applications
(EPR-Bell type experiments, quantum cryptography, quantum teleportation,
etc.). This is possible as far as the entanglement is not obscured by a
mixed state of the total system. Therefore, other measures have
been proposed to characterize the {\it operationally available}
entanglement in mixtures (Peres 1996, Vedral {\it et al.} 1997). However,
these measures do not represent the true and complete entanglement, since
a mixed state, which reduces this measure, is either
based on entanglement itself (on that of the whole bipartite system with
its environment), or the consequence of averaging over an ensemble of
unknown (but nonetheless entangled) states.

The eigenbasis of the reduced density matrix  can be used, by
othogonalizing the correlated ``relative states'' of the environment, to
write the total state as a single sum,
\be
      \ket\Psi = \sum_k \sqrt{p_k} \ket{\hat\phi_k^{\rm system}}
\ket{\hat\phi_k^{\rm environment}} \quad .
      \ee
While Erhard Schmidt (1907) first introduced this representation as a
mathematical theorem, Schr\"odinger (1935) used it for discussing
entanglement as representing what he called ``probability relations
between separated systems''.  It was later shown to be useful for
describing quantum nonlocality and decoherence by means of a universal
wave function (Zeh 1971, 1973).

The two orthogonal systems
$\hat\phi_k$ of the Schmidt form (2.16)  are {\it determined} (up to
degeneracy of the $p_k$'s) by the total state
$|  \Psi\rangle$ itself.
A time dependence of $| \Psi \rangle$
must therefore affect both the Schmidt states  {\it and} their (formal)
probabilities
$p_k$ (see K\"ubler and Zeh 1973, Pearle 1979, Albrecht 1993), hence also
the subsystem entropy, such as the linear entropy $\sum p_k (1 - p_k)$.
The induced subsystem dynamics is thus not ``autonomous". Similar to
the  motion of a shadow that merely reflects the regular motion of a
physical object, the reduced information content of the subsystem density
matrix by itself is insufficient to determine its own change. Likewise,
Boltzmann had to introduce his {\it Sto{\ss}zahlansatz}, based on
statistical assumptions, when neglecting {\it statistical} correlations
between particles (instead of quantum entanglement). The exact dynamics of
any  local ``system" would in general require the whole Universe to be
taken into account.

Effective ``open systems
quantum dynamics" has indeed been {\it postulated} in analogy to the
Boltzmann equation  by means of  semigroups or master equations for
calculating forward in time. An equivalent formalism was
introduced by Feynman and Vernon  (1963) in terms of path integrals.  As
explained above, this description can neither be exact, nor would it
justify the replacement of improper mixtures by proper ones {\it unless
explicitly postulated as a fundamental correction to the Schr\"odinger
equation}.  The formal theory of master equations will be discussed in
Chap.\ts 7 (see also Zeh 2001). Its foundation for local systems in
terms of a global unitary Schr\"odinger equation requires very specific
(statistically improbable) cosmic initial conditions.

According to a universal Schr\"odinger equation, quantum correlations with
the environment are permanently created with great efficiency for all
macroscopic systems, thus leading to decoherence, defined as the
irreversible dislocalization of phase relations (see Chap.\ts 3 for many
examples). The apparent (or ``improper") ensembles, obtained for
subsystems in this way, often led to  claims that decoherence be
able (or meant) to solve the measurement problem.\footnote{%
In the Schmidt basis,  interference
terms  are {\it exactly} absent by definition. Hepp (1972) used the
formal limit $N \to \infty $ to obtain this result in a {\it given} basis
(while this may require infinite time). However, the global state
always remains one pure superposition. The Schmidt representation has
therefore been used instead to specify the Everett  branching,
that is, to define the ultimate ``pointer basis"
$| \chi^{\rm observer}_n
\rangle$ for each observer (cf. Zeh 1973, 1979, Albrecht 1992, 1993,
Barvinsky and Kamenshchik 1995). It is also used in the ``modal
interpretation" of quantum mechanics (cf. Dieks 1995).}
   The apparent nature of these ensembles has then in turn been used to
declare the program of decoherence a failure. As explained in Sect.~2.3,
both claims miss the point. However, decoherence represents a {\it
crucial dynamical step} in the measurement process. The rest may
remain a pure epistemological problem (requiring only a reformulation of
the psycho-physical parallelism in quantum mechanical terms). If the
Schr\"odinger equation is exact, the observed quantum
indeterminism can only reflect that of the observer's identity -- not one
to be found in objective dynamics.

The process of decoherence leads to a novel, dynamically consistent
concept of a generalized course graining. If phase relations
between certain subspaces of a system permanently disappear by
decoherence, their reduced density matrices may be approximated in the
form
\be
      \rho
= \sum_{m,n}P_m \rho P_n
	    \approx  \sum_n P_n \rho P_n \quad ,
      \ee
where $P_n$ projects on to the $n$-th decohered subspace, while $\sum P_n
=
\bbbone$. The dynamics of the formal probabilities
$p_n(t) := \rm{tr} \{P_n \rho (t)\}$  may then be written as a master
equation, similar to the Pauli equation,
\be
      \dot p_n = \sum_m A_{nm} (p_m - p_n) \quad ,
      \ee
as was shown by Joos  (1984). Since (2.18)
describes  stochastic subsystem dynamics (in the direction of
time that is characterized by the process of decoherence), it defines
probabilities for coarse-grained ``histories"
$n(t)$, corresponding to time-ordered sequences of projections
$P_{n_1}(t_1)\ldots P_{n_k}(t_k)$. Probabilities for such
histories in discrete time steps can be written as
\be
p(n_1,\dots,n_k) = \rm{tr}\{P_{n_k}(t_k)\ldots P_{n_1}(t_1)\rho (t_0)\}
     \ee
(using of the property $P^2 = P$ of projection operators). States
$n_k$ dynamically arising according to a master equation may contain
``consistent memories" (or ``time capsules" in Barbour's words), while the
corresponding apparent histories appear ``quasi-classical" (robust under
decoherence).  Such histories would individually obey a
{\it quantum Langevin} equation (an indeterministic generalization of the
Schr\"odinger equation). Models (often assumed to hold {\it exactly}
instead of the Schr\"odinger equation) have been proposed  by Di\'osi
(1986), Belavkin (1988), Gisin and Percival (1992), and others -- see also
Di\'osi and Kiefer (2001).

In the theory of  ``consistent histories"  (Griffiths 1984,
Omn\`es 1992, 1995), {\it formal} projections $P_n$ are called
``events" regardless of any dynamics. These events are thus
{\it not} dynamically described within the theory -- in accord with the
Copenhagen interpretation, where events are assumed to occur ``out of
the blue" or ``outside the laws of nature". However, only those histories
$n_1,\dots n_k$ are then {\it admitted by postulate} (that is, assumed to
``occur") which possess ``consistent'' probabilities -- in the dynamical
sense of being compatible with a stochastic {\it evolution}. This
condition {\it requires} (a weak form of) decoherence, which is {\it not}
generally based on entanglement (cf. Omn\`es 1999).  This
dynamical dilemma is then resolved by  Griffiths and Omn\`es by
introducing a ``new logic". It culminates in Omn\`es' (1995) surprising
remark that  ``the formalism of  logic is not time-reversal invariant, as
can be seen in the time ordering of the (projection) operators".
However, a property
$a$ at time
$t_1$ that is said to ``imply" a property $b$ at time  $t_2 > t_1$
would describe a {\it causal} (that  is, dynamical) rather than a {\it
logical} relationship.  This conceptual confusion of cause and reason
seems to have a  long tradition in philosophy, while even in mathematics
the truth of logical theorems is often inappropriately {\it defined} by
means of logical {\it operations} that have to be performed in time
(thus mimicking a causal relation).

In the theory of decoherence, {\it apparent} events in the detector are
described dynamically by the universal Schr\"odinger equation, using
certain initial conditions for the environment, as a very fast but smooth
formation of entanglement. Similarly,
``particles" {\it appear} in the form of narrow wave packets in space as
a consequence of decoherence in the detector. This
identification of observable events with a decoherence process holds
regardless of any conceivable subsequent genuine collapse.
Therefore, {\it decoherence is not only responsible for the
classical aspects of quantum theory, but also for its ``quantum" aspects}
(see Sect.~3.3.2.3). All fundamental physical concepts are  continuous and
based on ``smooth" Schr\"odinger dynamics.

This
description of quantum events as a unitary process also avoids  any
``superluminal effects" that have been shown (with mathematical rigour --
see Hegerfeldt 1994)
   to arise (not very surprisingly) from
explicitly or tacitly assumed instantaneous quantum  jumps between exact
energy or particle  number eigenstates.\footnote{%
Such superluminal ``phenomena'' are reminiscent of the story of {\it Der
Hase und der Igel} (the race between {\it The Hedgehog and the Rabbit}),
narrated by the Grimm brothers. Here, the hedgehog, as a competitor in
the race, does not run at all, while his wife is waiting at the end of
the furrow, shouting in low German ``Ick bin all hier!" (``I'm already
here!''). Similar arguments hold for ``quantum teleportation" (cf.~also
Vaidman 1998). Experiments clearly support the view that reality is
described by a smoothly evolving wave function, nonlocal but dynamically
compatible with relativity, rather than in terms of probabilistic
``events'' (cf.~Fearn, Cock, and Milonni 1995). It is the local observer
whose identity ``splits'' indeterministically according to the
Schr\"odinger equation. Superluminal teleportation {\it would} be
required to describe the corresponding experiments if (local) physical
properties entered existence ``out of the blue'' in
fundamental quantum events.}   The latter require infinite exponential
tails that can never {\it form} completely in a relativistic world.
Supporters of explicit collapse mechanisms are quite aware of this
problem, and try to avoid it (cf. Di\'osi and Luk\'acz 1994 and Chap.~8).
In the nonlocal quantum formalism, {\it dynamical locality} is achieved
by using Hamiltonian operators that are spatial integrals over a
Hamiltonian density. This form prevents superluminal signalling and the
like.

\section    {Conclusion}

Let me recall the interpretation of quantum theory that has now emerged in
accordance with the concept of decoherence:

\noindent (1) General quantum superpositions (such as a wave function)
represent individual physical states (Sect.\ts 2.1.1).

\noindent (2) According to a universal Schr\"odinger
equation, most superpositions are almost immediately, and in practice
irreversibly, {\it dislocalized} by interaction with their environment.
Although the resulting nonlocal superpositions still {\it exist}, we do
not know, in general, what they {\it mean} (or how they could be
observed). However, if dynamics is local (described by a Hamiltonian
density in space, $H=\int h({\bf r}) d^3
\bf r)$, approximately factorizing components may be dynamically
autonomous after this decoherence has occurred, and nonlocal
superpositions cannot return into local ones if statistical arguments
apply to the future (Zeh 2001).

\noindent (3) Any observer (assumed to be local for empirical and
dynamical reasons) who attempts to observe a subsystem of the nonlocal
superposition must become part of this entanglement. Those of his
component states which are then related only by  nonlocal phase
relations describe different observations. So we may
axiomatically identify these individual {\it component} states of the
observer with states of consciousness (novel psycho-physical
parallelism).

\noindent (4) Because of this dynamical autonomy of decohered
world components (``branches''), there is no reason to deny the existence
of  ``the other'' components which result from the Schr\"odinger equation
(``many minds interpretation'' -- Sect.~2.3).

\noindent (5) Probabilities are meaningful only as frequencies in series
of repeated measurements. In order to derive the observed Born
probabilities in terms of frequencies, we have to {\it postulate} merely
that we are living in an ``Everett branch'' with not extremely small norm.
 

\section*{References}
\begin{list}{}
{
\setlength{\itemsep}{0.2cm}
\setlength{\leftmargin}{0.3cm}
\setlength{\itemindent}{-0.3cm}
}
\item
Aharonov, Y., and Vaidman, L. (1991): Complete description of a
quantum system at a given time. {\it J. Phys.} {\bf A24}, 2315.
\item
Albert, D. and Loewer, M. (1988): ``Interpreting the Many Worlds
Interpretation'' {\it Synthese}  {\bf 77}, 195.
\item
Albrecht, A. (1992): ``Investigating decoherence in a simple system.''
{\it Phys. Rev.}  {\bf D46}, 5504.
\item
Albrecht, A. (1993): ``Following a `collapsing' wave function.'' {\it
Phys. Rev.}  {\bf D48}, 3768.
\item
Arndt, M., Nairz, O., Vos-Andreae, J., Keler, C., van der Zouw, G.,
and Zeilinger, A. (1999):  Wave-particle duality of $C_{60}$
molecules. {\it Nature} {\bf 401}, 680.
\item
Barbour, J.B. (1994): ``The timelessness of quantum gravity: II. The
appearance of dynamics in static configurations.'' {\it Class. Quantum
Grav.} {\bf 11}, 2875.
\item
Barbour, J.B. (1999): {\it The End of Time} (Weidenfeld \& Nicolson,
London).
\item
Bardeen, J., Cooper, L.N., and Schrieffer, J.R. (1957): ``Theory of
superconductivity.'' {\it Phys. Rev.}  {\bf 108}, 1175.
\item
Barvinsky, A.O. and Kamenshchik, A.Yu. (1995): ``Preferred basis
in quantum theory and the problem of classicalization of the quantum
Universe.'' {\it Phys. Rev.}  {\bf D52}, 743.
\item
Belavkin, V.P. (1988): ``Nondemolition measurements, nonlinear
filtering and dynamic programming of quantum stochastic processes."
In: {\it Modeling and Control of Systems. Lect. Notes in Control and Inf.
Sciences} {\bf 121}, ed. by A. Blaqui\'ere (Springer, Berlin), p 245.
\item
Bell, J.S. (1964): ``On the Einstein Podolsky Rosen paradox.'' {\it
Physics}  {\bf 1}, 195 -- reprinted in Wheeler and Zurek (1983) and in
Bell (1987).
\item
Bell, J.S. (1981): ``Quantum mechanics for cosmologists.'' In: {\it
Quantum Gravity 2}, ed. by C. Isham, R. Penrose and D. Sciama
(Clarendon Press, Oxford), p. 611 -- reprinted in Bell (1987).
\item
Bell, J.S. (1987): {\it Speakable and Unspeakable in Quantum
Mechanics}  (Cambridge University Press).
\item
Bennett, C.H. (1973): ``Logical reversibility of computation.'' {\it
IBM J. Res. Dev.}  {\bf 17}, 525.
\item
Bertet, P., Osnaghi, S., Rauschenbeutel, A., Nogues, G., Auffeves, A.,
Brune, M., Raimond, J.M., and Haroche, S. (2001): ``A complementarity
experiment with an interferometer at the quantum-classical boundary.''
Nature {\bf 441}, 166.
\item
Bohm, D. (1952): ``A Suggested Interpretation of the Quantum
Theory in Terms of `Hidden' Variables.'' {\it Phys. Rev.}  {\bf 85},
166 -- reprinted in Wheeler and Zurek (1983).
\item
Bohr, N. (1928): ``The Quantum Postulate and the Recent
Development of Atomic Theory.'' {\it Nature}  {\bf 121}, 580 -- reprinted
in Wheeler and Zurek (1983).
\item
Bollinger, J.J., Heinzen, D.J., Itano, W.M., Gilbert, S.L., and
Wineland, D.J. (1989): ``Test of the linearity of quantum mechanics by
rf spectroscopy on the $^9$Be$^+$ ground state.'' {\it Phys. Rev. Lett.}
{\bf 63}, 1031.
\item
Borel, E. (1914): {\it Le Hasard}  (Alcan, Paris).
\item
Born, M. (1926): ``Das Adiabatenprinzip in der Quantenmechanik.''
{\it Z. Phys.} {\bf 40}, 167.
\item
Born, M. (1969): {\it Albert Einstein, Hedwig und Max Born,
Briefwechsel 1916}  (Nymphenburger Verlagshandlung,
M\"unchen).
\item
Brillouin, L. (1962): {\it Science and Information Theory}
(Academic Press, New York).
\item
Brukner, C., and Zeilinger, A. (2000): ``Conceptual Inadequacy of the
Shanon Information in Quantum Measurements." Report quant-ph/0006087.
\item
Brune, M., Hagley, E., Dreyer, J., Maitre, X., Moali, Y., Wunderlich, C.,
Raimond, J.M., and Haroche, S. (1996): ``Observing the Progressive
Decoherence of the `Meter' in a Quantum Measurement.'' {\it Phys. Rev.
Lett.}  {\bf 77}, 4887.
\item
Busch, P., Cassenelli, G., De Vito, E., Lahti, P., and Levrero, A.
(2001): ``Teleportation and Measurements.'' {\it Phys. Lett.} {\bf A284},
141.
   %
\item
Caldeira, A.O. and Leggett, A.J. (1983): ``Path Integral Approach
to Quantum Brownian Motion.'' {\it Physica}
    {\bf 121A}, 587.
\item
d'Espagnat, B. (1966): ``An elementary note about mixtures.''. In:
{\it Preludes in theoretical physics}, ed. by De-Shalit, Feshbach and v.
Hove (North-Holland Publishing Comp., Amsterdam), p. 185.
\item
d'Espagnat, B. (1976): {\it Conceptual Foundations of Quantum
Mechanics}  (W. A. Benjamin, Reading, MA).
\item
d'Espagnat, B. (1995): {\it Veiled Reality}  (Addison-Wesley,
Reading/MA.).
\item
d'Espagnat, B. (2001): ``A note on measurement."
{\it Phys. Lett.} {\bf A282}, 133 -- quant-ph/0101141.
   %
\item
Dieks, D. (1995): ``Physical motivation of the modal interpretation of
quantum mechanics.'' {\it Phys. Lett.}  {\bf A197}, 367.
\item
Di\'osi, L. (1985): ``Orthogonal Jumps of the Wavefunction in
White-Noise Potentials.'' {\it Phys. Lett.}  {\bf A112}, 288.
\item
Di\'osi, L. (1986): ``Stochastic pure state representation for open
quantum systems.'' {\it Phys. Lett.}  {\bf A114}, 451.
\item
Di\'osi, L. and Kiefer C. (2000): ``Robustness and Diffusion of Pointer
States.'' {\it Phys. Rev. Lett.} {\bf 85}, 3552.
\item
Di\'osi, L. and Luk\'acz, B. (1994): {\it Stochastic Evolution of
Quantum States in Open Systems and in Measurement Processes}  (World
Scientific, Singapore).
\item
Dirac, P.A.M. (1947): {\it The Principles of Quantum Mechanics}
(Clarendon Press, Oxford).
\item
Donald, M.J. (1995): ``A Mathematical Characterization of the
Physical Structure of Observers.'' {\it Found. Phys.}  {\bf 25}, 529.
\item
Dyson, F. (1949): ``The S Matrix in Quantum Electrodynamics.", {\it Phys.
Rev.} {\bf 75}, 1736.
\item
Ellis, J., Mohanty, S., and Nanopoulos, D.V. (1989): ``Quantum
Gravity and the Collapse of the Wave Function.'' {\it Phys. Lett.}  {\bf
B221}, 113.
\item
Englert, B.G., Scully, M.O., S\"ussmann, G., and Walther, H. (1992):
``Surrealistic Bohm trajectories." {\it Z. Naturf.} {\bf 47a}, 1175.
\item
Everett, H., III. (1957): `` `Relative State' Formulation of Quantum
Mechanics.'' {\it Rev. Mod. Phys.}  {\bf 29}, 454 -- reprinted in Wheeler
and Zurek (1983).
\item
Fearn, H., Cook, R.J., and Milonni, P.W. (1995): ``Sudden replacement
of a Mirror by a Detector in Cavity QED: Are Photons Counted Immediately?"
{\it Phys. Rev. Lett.} {\bf 74}, 1327.
\item
Feynman, R.P., Leighton, R.B., and Sands, M. (1965): {\it The
Feynman Lectures on Physics}  (Addison-Wesley, Reading).
\item
Feynman, R.P. and Vernon, F.L., jr. (1963): ``The Theory of a
General Quantum System Interacting with a Linear Dissipative System.''
{\it Ann. Phys. (N.Y.)}  {\bf 24}, 118.
\item
Friedman, J.R., Vijay, P.W., Chen, S.K., Tolpygo, S.K., and Lukens, J.E.
(2000): ``Quantum superposition of distinct macroscopic states.'' {\it
Nature} {\bf 406}, 43.
\item
Gell-Mann, M. and Pais, A. (1955): ``Behavior of Neutral Particles
under Charge Conjugation.'' {\it Phys. Rev.}  {\bf 97}, 1387.
\item
Ghirardi, G.C., Rimini, A., and Weber, T. (1986): ``Unified
dynamics for microscopic and macroscopic systems.'' {\it Phys. Rev.}
{\bf D34}, 470.
\item
Gisin, N. and Percival, I.C. (1993): ``Quantum state diffusion,
localization and quantum dispersion entropy". {\it J. Phys. A: Math.
Gen.} {\bf 26}, 2233
\item
Giulini, D., Kiefer, C., and Zeh, H.D. (1995): ``Symmetries,
superselection rules, and decoherence.'' {\it Phys. Lett.}  {\bf A199},
291.
\item
Glauber, R.J. (1963): ``Coherent and Incoherent States of the
Radiation Field.'' {\it Phys. Rev.}  {\bf 131}, 2766.
\item
Graham, N. (1970): {\it The Everett Interpretation of Quantum
Mechanics}  (Univ. North Carolina, Chapel Hill).
\item
Griffiths, R.B. (1984): ``Consistent Histories and the Interpretation
of Quantum Mechanics.'' {\it J. Stat. Phys.}
    {\bf 36}, 219.
\item
Gross, D.J. (1995): ``Symmetry in Physics: Wigner's Legacy."
{\it Phys. Today} {\bf 48}, (Dec.) 46.
\item
Haag, R. (1992): {\it Local Quantum Physics} (Springer, Berlin).
\item
Hawking, S.W. (1987): ``Quantum Coherence down the Wormhole." {\it Phys.
Lett.} {\bf 195B}, 337.
\item
Hegerfeldt, G. (1994): ``Causality Problems for Fermi's Two-Atom
System.'' {\it Phys. Rev. Lett.}  {\bf 72}, 596.
\item
Hepp, K. (1972): ``Quantum theory of measurement and macroscopic
observables.'' {\it Helv. Phys. Acta} {\bf 45}, 236.
\item
Herzog, Th.J., Kwiat, P.G., Weinfurter, H., and Zeilinger, A. (1995):
``Complementarity and The Quantum Eraser." {\it Phys. Rev. Lett.} {\bf
75}, 3034.
\item
Hund, F. (1927): ``Zur Deutung der Molekelspektren. III.'' {\it Z.
Phys.} {\bf 43}, 805.
\item
Jauch, J.M. (1968): {\it Foundations of Quantum Mechanics}
(Addison-Wesley, Reading, MA).
\item
Joos, E. (1984): ``Continuous Measurement: Watchdog Effect versus
Golden Rule.'' {\it Phys. Rev.}  {\bf D29}, 1626.
\item
Joos, E. (1986): ``Why do we observe a classical spacetime?'' {\it
Phys. Lett.}  {\bf A116}, 6.
\item
Joos, E. and Zeh, H.D. (1985): ``The Emergence of Classical
Properties Through Interaction with the Environment.'' {\it Z. Phys.}
{\bf B59}, 223.
\item
Jordan, P. and Klein, O. (1927): ``Zum Mehrk\"orperproblem in der
Quantentheorie.'' {\it Z. Phys.}  {\bf 45}, 751.
\item
Kiefer, C. (1987): ``Continuous measurement of mini-superspace
variables by higher multipoles.'' {\it Class. Quantum Grav.}  {\bf 4},
1369.
\item
Kiefer, C. (1992): ``Decoherence in quantum electrodynamics and
quantum cosmology.'' {\it Phys. Rev.}
    {\bf D46}, 1658.
\item
K\"ubler, O. and Zeh, H.D. (1973): ``Dynamics of Quantum
Correlations.'' {\it Ann. Phys. (N.Y.)}  {\bf 76}, 405.
\item
Lee, T.D. and Yang, C.N. (1956): ``Question of Parity Conservation
in Weak Interactions.'' {\it Phys. Rev.}  {\bf 104}, 254.
\item
Lockwood, M. (1989): {\it Mind, Brain and the Quantum: The
Compound 'I'}  (Basil Blackwood, Oxford).
\item
London, F. and Bauer, E. (1939): {\it La th\'eorie d'observation en
M\'ecanique Quantique}  (Hermann, Paris); English
translation in Wheeler and Zurek (1983).
\item
L\"uders, G. (1951): ``\"Uber die Zustands\"anderung durch den
Me{\ss}proze{\ss}.'' {\it Ann. Phys. (Leipzig)}  {\bf 8}, 322.
\item
Ludwig, G. (1990): `{\it Die Grundlagen einer physikalischen Theorie}
(Springer, Berlin).
\item
Machida, S. and Namiki, M. (1980): ``Theory of Measurement in
Quantum Mechanics.'' {\it Progr. Theor. Phys.}  {\bf 63}, 1457.
\item
Mensky, M.B. (1979): ``Quantum restrictions for continuous
observation of an oscillator.'' {\it Phys. Rev.}  {\bf D20}, 384.
\item
Mensky, M.B. (2000): {\it Quantum Measurements and Decoherence}
(Kluwer Academic).
\item
Mirman, R. (1970): ``Analysis of the Experimental Meaning of
Coherent Superposition and the Nonexistence of Superselection Rules.''
{\it Phys. Rev.}  {\bf D1}, 3349.
\item
Monroe, C., Meekhof, D.M., King, B.E., and Wineland, D.J. (1996):
``A `Schr\"o\-dinger Cat' Superposition State of an Atom." {\it Science}
{\bf 272}, 1131.
\item
Mooij, J.E., Orlando, T.P., Levitov, L., Tian, L., van der Wal, C.H., and
Lloyd, S. (1999): ``Josephson Persistent-Current Qubit'' {\it Science}
{\bf 285}, 1036.
\item
Omn\`es, R. (1992): ``Consistent interpretation of quantum
mechanics.'' {\it Rev. Mod. Phys.}  {\bf 64}, 339.
\item
Omn\`es, R. (1995): ``A New Interpretation of Quantum Mechanics
and Its Consequences in Epistemology.'' {\it Found. Phys.}  {\bf 25},
605.
\item
Omn\`es, R. (1999):  {\it Understanding Quantum Mechanics}
(Princeton UP).
\item
Page, D.N. (1995): ``Sensible quantum mechanics: are only
perceptions probabilistic?'' Report quant-ph/9506010.
\item
Pearle, P. (1976): ``Reduction of the state vector by a nonlinear
Schr\"odinger equation.'' {\it Phys. Rev.}  {\bf D13}, 857.
\item
Pearle, P. (1979): ``Toward Explaining Why Events Occur.'' {\it Int.
J. Theor. Phys.}  {\bf 18}, 489.
\item
Pearle, P. and Squires, E. (1994): ``Bound State Excitation, Nucleon
Decay Experiments, and Models of Wave Function Collapse.'' {\it Phys.
Rev. Lett.}  {\bf 73},~1.
\item
Peres, A. (1996): Separability Criterion for Density Matrices. Phys.
{\it Rev. Lett.} {\bf 77}, 1413.
\item
Primas, H. (1981): {\it Chemistry, Quantum Mechanics and
Reductionism} (Springer, Ber\-lin).
\item
Primas, H. (1990): ``Mathematical and philosophical questions
in the theory of open and macroscopic quantum systems." In: {\it
Sixty-Two Years of Uncertainty.}, ed. by A.I. Miller (Plenum Press,
New York), p. 233.
\item
Rauch, H., Zeilinger, A., Badurek, G., Wilfing, A., Bauspiess, W.,
and Bonse, U. (1975): ``Verification of Coherent Spinor Rotations of
Fermions.'' {\it Phys. Lett.}  {\bf 54A}, 425.
\item
Rempe, G., Walther, H., and Klein, N. (1987): ``Observation of
Quantum Collapse and Revival in a One-Atom Maser.'' {\it Phys. Rev.
Lett.} {\bf 58}, 353.
\item
Rhim, W.-K., Pines, A., and Waugh, J.S. (1971): ``Time-Reversal
Experiments in Dipolar-Coupled Spin Systems.'' {\it Phys. Rev.}  {\bf
B3}, 684.
\item
Schmidt, E. (1907): ``Zur Theorie der linearen und nichtlinearen
Integralgleichungen. Zweite Abhandlung: Aufl\"osung der allgemeinen
linearen Integralgleichung.''
   {\it Math. Annalen}  {\bf 64}, 161.
\item
Schr\"odinger, E. (1935): ``Discussion of probability relations
between separated systems.'' {\it Proc. Cambridge Phil. Soc.}  {\bf 31},
555.
\item
Shi, Y. (2000): ``Early gedanken experiments of quantum mechanics
revisited.'' Annalen Phys. {\bf 9}, 637 - quant-ph/9811050.
\item
Squires, E. (1990): {\it Conscious Mind in the Physical World}
(Hilger, Bristol).
\item
Srinivas, M.D. (1984): ``Quantum theory of continuous
measurements.'' In: {\it Quantum Probability and Applications}, ed. by
L. Accardi  {\it et al.} (Springer, Berlin).
\item
Stapp, H.P. (1993): {\it Mind, Matter, and Quantum Mechanics}
(Springer, Berlin).
\item
Streater and Wightman (1964): {\it TCP, Spin and Statistics, and All
That}  (Addison-Wesley, Reading).
\item
Szilard, L. (1929): ``\"Uber die Entropieverminderung in einem
thermodynamischen Sy\-stem bei Eingriffen intelligenter Wesen.'' {\it Z.
Physik\/}  {\bf 53}, 840--856; English translation
   in Wheeler and Zurek (1983).
\item
Tegmark, M. (2000): ``Importance of quantum decoherence in brain
processes." {\it Phys. Rev.} {\bf 61}, 4194.
\item
Tegmark, M., and Wheeler, J.A. (2001): ``100 Years of Quantum
Mysteries." {\it Scientific American} {\bf 284} (February), 54 --
quant-ph/0101077.
\item
Tessieri, L., Vitali, D., and Grigolini, P. (1995): ``Quantum
   jumps as an objective process of nature.'' {\it Phys. Rev.}
   {\bf A51}, 4404.
\item
Tittel, W., Brendel, T., Zbinden, H., and Gisin,
N. (1998):  Violation of Bell Inequalities by Photons More Than 10
km Apart. {\it Phys. Rev. Lett.} {\bf 81}, 3563.
\item
Ulfbeck, O., and Bohr, A. (2001): ``Genuine fortuitousness: Where
did that click come from." {\it Found. Phys.} {\bf 31}, 757.
\item
Vaidman, L. (1998): ``Teleportation: Dream or Reality.'' Report
quant-ph/9810089.
\item
Vedral, V., Plenio, M.B., Rippin, M.A., Knight, P.L. (1997):
``Quantifying Entanglement.`` {\it Phys. Rev. Lett.} {\bf 78}, 2275.
\item
Venugopalan, A., Kumar, D., and Gosh, R. (1995):
``Environment-induced decoherence II. Effect of decoherence on
Bell's inequality for an EPR pair."
{\it Physica} {\bf A220}, 576.
\item
von Neumann, J. (1932): {\it Mathematische Grundlagen der
Quantenmechanik}  (Springer, Berlin); reprinted 1981. English
   translation by
R.T. Beyer (1955): {\it Mathematical Foundations of Quantum
Mechanics} (Princeton University Press).
\item
Weinberg, S. (1989): ``Precision Tests of Quantum Mechanics.'' {\it
Phys. Rev. Lett.}  {\bf 62}, 485.
\item
Wess, J. and Zumino, B. (1971): ``Consequences of anomalous Ward
identities.'' {\it Phys. Lett.}  {\bf 37B}, 95.
\item
Wheeler, J.A. and Zurek, W.H. (1983): {\it Quantum Theory and
Measurement}  (Princeton University Press).
\item
Wick, G.C., Wightman, A.S., and Wigner, E.P. (1970):
``Superselection Rule for Charge.'' {\it Phys. Rev.}  {\bf D1},
   3267.
\item
Wightman, A.S. (1995): ``Superselection Rules; Old and New.'' {\it
Il Nuovo Cimento}  {\bf 110B}, 751.
\item
Wigner, E.P. (1962): ``Remarks on the Mind-Body Question.'' In:
{\it The Scientist Speculates}, ed. by L.G. Good (Heinemann, London),
p. 284 -- reprinted in Wheeler and Zurek (1983).
\item
Wigner, E.P. (1964): ``Events, Laws of Nature, and Invariance
Principles.'' In: {\it Les Prix Nobel en 1963} (The Noble Foundation,
Stockholm).
\item
Wolfenstein, L. (1978): ``Neutrino oscillations in matter.'' {\it Phys.
Rev.}  {\bf D17}, 2369.
\item
Woolley, R.G. (1986): ``Molecular Shapes and Molecular
Structures.'' {\it Chem. Phys. Lett.}  {\bf 125}, 200.
\item
Yuen, H.P. (1976): ``Two-photon coherent states of the radiation
field.'' {\it Phys. Rev.}  {\bf A13}, 2226.
\item
Zeh, H.D. (1970): ``On the Interpretation of Measurement in
Quantum Theory.'' {\it Found. Phys.}  {\bf 1}, 69
   -- reprinted in Wheeler and Zurek (1983).
\item
Zeh, H.D. (1971): ``On the Irreversibility of Time and Observation
in Quantum Theory.'' In: {\it Foundations of Quantum Mechanics}, Varenna
1970, ed. by B. d'Espagnat (Academic Press, NewYork),
p.\ts 263.
\item
Zeh, H.D. (1973): ``Toward a Quantum Theory of Observation.''
{\it Found. Phys.}  {\bf 3}, 109.
\item
Zeh, H.D. (1979): ``Quantum Theory and Time Asymmetry." {\it Found. Phys.}
{\bf 9}, 803.
\item
Zeh, H.D. (1981): ``The Problem of Conscious Observation in
Quantum Mechanical Description.'' {\it Epist. Letters}
   {\bf 63.0} (Ferd. Gonseth Ass., Biel) -- see also Zeh (2000).
\item
Zeh, H.D. (1986): ``Emergence of Classical Time from a Universal
Wave Function.'' {\it Phys. Lett.}  {\bf A116}, 9.
\item
Zeh, H.D. (1988): ``Time in Quantum Gravity.'' {\it Phys. Lett.}
{\bf A126}, 311.
\item
Zeh, H.D. (1999): ``Why Bohm's Quantum Theory?'' {\it Found. Phys. Lett.}
{\bf 12}, 197 -- quant-ph/9812059.
\item
Zeh, H.D. (2000): ``The Problem of Conscious Observation in Quantum
Mechanical Description." {\it Found. Phys. Lett.} {\bf 13}, 221 --
quant-ph/9908084.
\item
Zeh, H.D. (2001): {\it The Physical Basis of the Direction
of Time}, 4$^{th}$ edn.   (Springer, Berlin) -- see also
www.time-direction.de.
\item
Zurek, W.H. (1981): ``Pointer basis of quantum apparatus: Into what
mixture does the wave packet collapse?'' {\it Phys. Rev.}  {\bf D24},
1516.
\item
Zurek, W.H. (1982): ``Environment-induced superselection rules.''
{\it Phys. Rev.}  {\bf D26}, 1862.
\item
Zurek, W.H. (1984): ``Maxwell's demon, Szilard's engine, and
quantum measurements.'' In: {\it Frontiers in Nonequilibrium Statistical
Mechanics}, ed. by G.R. Moore and M.O. Scully (Plenum, New York).
\item
Zurek, W.H. (1991): ``Decoherence and the Transition from
Quantum to Classical.'' {\it Physics Today}  {\bf 44} (Oct.), 36.
\item
Zurek. W.H. (2001): ``Decoherence, Einselection, and the Quantum Origin
of the Classical." {\it Rev. Mod. Phys.} (to be published) --
quant-ph/0105127.
\item
Zurek, W.H., Habib, S., and Paz, J.P. (1993): ``Coherent States via
Decoherence.'' {\it Phys. Rev. Lett.}  {\bf 70}, 1187.
\item
Zurek, W.H. and Paz, J.P. (1994): ``Decoherence, Chaos and the
Second Law.'' {\it Phys. Rev. Lett.}  {\bf 72}, 2508.
\end{list}

\end{document}